\documentclass[11pt,review]{article}
\usepackage[a4paper,margin=0.787in]{geometry}
\usepackage{amsmath}
\usepackage{amsfonts}
\usepackage{amssymb}
\usepackage{graphicx}
\usepackage[colorlinks=true, allcolors=blue]{hyperref}
\usepackage{pdfpages}
\usepackage{enumitem}
\usepackage{setspace}
\usepackage{xcolor}
\usepackage{authblk}
\usepackage{algorithm}
\usepackage{algpseudocode}
\usepackage{fullpage}
\usepackage[numbers,sort&compress]{natbib}
\usepackage{multirow}
\newcommand{\YY}[1]{#1}

\title{The influence of cell phenotype on collective cell \\ invasion into the extracellular matrix}
\author[1]{Yuan Yin}
\author[1]{Sarah L.~Waters}
\author[1]{Ruth E.~Baker}
\affil[1]{Mathematical Institute, University of Oxford, Oxford, UK}
\date{}

\begin{document}
\maketitle


\begin{abstract}
Understanding the interactions between cells and the extracellular matrix (ECM) during collective cell invasion is crucial for advancements in tissue engineering, cancer therapies, and regenerative medicine. This study focuses on the roles of contact guidance and ECM remodelling in directing cell behaviour, with a particular emphasis on exploring how differences in cell phenotype impact collective cell invasion. We present a computationally tractable two-dimensional hybrid model of collective cell migration within the ECM, where cells are modelled as individual entities and collagen fibres as a continuous tensorial field. Our model incorporates random motility, contact guidance, cell-cell adhesion, volume filling, and the dynamic remodelling of collagen fibres through cellular secretion and degradation. Through a comprehensive parameter sweep, we provide valuable insights into how differences in the cell phenotype, in terms of the ability of the cell to migrate, secrete, degrade, and respond to contact guidance cues from the ECM, impacts the characteristics of collective cell invasion. 
\end{abstract}


\section{Introduction}
\label{sec:introduction}

Collective cell invasion is a fundamental, tightly coordinated process involving both migration and proliferation, and it takes place throughout development, wound healing, and cancer metastasis~\cite{friedl2009collective,ilina2009mechanisms,mayor2016front,stock2021self}. This coordinated invasion is driven by cell--cell interactions and extrinsic cues provided by the surrounding microenvironment, with the extracellular matrix (ECM) playing a pivotal role~\cite{yue2014biology}. The ECM is a complex and dynamic macromolecular network composed of various proteins, such as collagen, fibronectin, and glycosaminoglycans~\cite{karamanos2021guide}. Collagen fibres are the predominant component of the ECM, and are essential for maintaining the mechanical integrity of tissues under stress~\cite{rezvani2021collagen}. Cells actively remodel the ECM by secreting, degrading, and reorganising its components as they migrate~\cite{frantz2010extracellular}. This dynamic interplay between cells and the ECM underlies many physiological and pathological processes, however its huge complexity means that it remains challenging to fully characterise using experiments alone. An additional complicating factor is that cells exhibit phenotypic heterogeneity, reflected in their response to their local biomechanical environment, metabolic activity, protrusive activity and connectivity patterns~\cite{Cheung2013,beunk2022actomyosin,agnihotri2015metabolic,haeger2015collective,Fried2017,Wolf2007}. 

Mathematical modelling offers a powerful approach to systematically investigate the roles of various cell--cell and cell--ECM interactions in driving collective cell migration, and to explore how differences in cell phenotype drive patterns of collective cell invasion. The overarching aim of this work is to develop a computationally tractable model of collective cell migration that includes both cell--cell interactions, such as adhesion and volume filling, and cell--ECM interactions, which lead to contact guidance and ECM \YY{evolution}, and use it to explore how they together drive collective cell migration. In particular we focus on elucidating how differences in the cell phenotype, in terms of their ability to migrate, secrete, degrade, and interact with the ECM, impact collective cell invasion signatures.


\subsection{Cell--ECM interactions during collective migration}

The ECM is not merely a passive mechanical support; it provides a dynamic environment that facilitates biochemical signalling, influencing cell activities such as migration, differentiation, and proliferation~\cite{yue2014biology}. Comprehensive reviews of ECM components, properties, and roles in cellular functions can be found in~\cite{kyriakopoulou2023trends,padhi2020ecm,walma2020extracellular}. Among the mechanisms by which the ECM influences cell motility is haptotaxis, where cells migrate along gradients of ECM-bound molecules. Haptotaxis a key process in tissue regeneration and cancer invasion~\cite{pekmen2024numerical,sfakianakis2020hybrid}. Another mechanism is durotaxis, where cells migrate towards regions of stiffer ECM, a process linked to cancer metastasis~\cite{espina2022durotaxis}. Finally, contact guidance is another important cell--ECM mechanism---here, collagen fibre alignment provides directional cues that reorients cell migration paths~\cite{pamonag2022individual,kim2021mechanics}. Experimental evidence demonstrates a key role for contact guidance in wound healing, where fibroblasts migrate by aligning with collagen fibres at the wound site~\cite{lackie2012cell}. There is also experimental evidence that contact guidance modulates the speed of cell migration~\cite{doyle2009one}.

The interaction between cells and the ECM is not unidirectional; cells actively remodel the ECM by secreting, degrading, and reorganising its components~\cite{frantz2010extracellular}. Fibroblasts, for example, can secrete collagen fibres that are aligned in the direction of their movement, thereby reinforcing pathways for future migration~\cite{mcdougall2006fibroblast}. Cells also employ matrix metalloproteinases (MMPs) to degrade ECM components, a process that is particularly well-studied in the context of cancer invasion and metastasis~\cite{reunanen2013matrix}. Additionally, cells interact mechanically with the ECM by exerting forces that remodel its structure. For instance, fibroblasts exert traction forces on newly formed granulation tissue, which increases its rigidity. This increasing stiffness strengthens cell--ECM contacts, encourages the formation of intracellular contractile stress fibres, drives fibroblast differentiation into proto-myofibroblasts, and facilitates migration~\cite{hinz2003mechanisms}. Recent work also highlights that fibroblasts exhibit mechanical memory when subjected to prolonged stretching: moderate stretching activates fibroblasts, while excessive stretching inhibits them~\cite{Weihs2025}.

The intricate \YY{bidirectional} feedback between cells and the ECM during collective cell invasion presents a significant challenge in disentangling how specific mechanisms drive certain dynamics and how these dynamics, in turn, influence each other. In this study, we aim to gain insights through the development of a \YY{tractable} mathematical model to address the overarching question of how the \YY{bidirectional} interactions between cells and the ECM drive the dynamics of collective cell migration while simultaneously shaping the architecture of the ECM, and how these patterns of collective invasion are altered in cell populations with different phenotypes.


\subsection{Mathematical models of ECM--cell interactions}

A key challenge in mathematical modelling of collective cell invasion is striking a balance between biological realism and computational tractability. Models must incorporate sufficient mechanistic detail to yield meaningful biological conclusions while remaining simple enough to allow for systematic parameter exploration. A broad range of modelling approaches have been developed to study cell--ECM interactions, as reviewed in~\cite{crossley2024modeling}. 

Of particular relevance to this study are several modelling frameworks that focus on the interplay between migrating cells and components of the ECM. Dallon et al.~\cite{dallon1999mathematical} models cells as discrete agents and collagen fibres as a continuous bidirectional vector field. The direction of cell migration is determined by two factors: the tendency of a cell to persist in its prior path, and the directional cues from the surrounding collagen fibres. To account for the presence of multiple collagen fibres at each point in space, Hillen~\cite{hillen2006m5} and Painter~\cite{painter2009modelling} introduced more complex models in which collagen fibres are represented by a probability distribution function that varies with orientation, space, and time. Cell movement is described by a velocity-jump process, formulated using a transport equation in which contact guidance depends on the surrounding collagen distributions. Though they capture many of the salient features of cell invasion into the ECM, both models neglect direct cell--cell interactions. 

Cumming et al.~\cite{cumming2010mathematical} extended the approach of Dallon et al.~\cite{dallon1999mathematical} to include additional complexity of the ECM, developing a six-species hybrid model that considers the dynamics of fibrin, collagen, macrophages, fibroblasts, transforming growth factor-$\beta$ (TGF-$\beta$) and tissue plasminogen activator. Collagen fibres dynamics are represented by a continuous tensorial field that captures fibre orientation and density, with computational tractability retained by decomposing the tensorial field into the sum of its (orthogonal) eigenvectors’ outer products. Each cell is represented as a circular disc, and its migration is influenced by mechanisms such as contact guidance and chemotaxis. Although it captures many features of the ECM, as well as cell signalling, the simple representation of cells in this model precludes study of how cell--cell interactions, such as cell--cell adhesion and volume filling, impact collective migration. 

Recent efforts have increased model complexity to study specific biological phenomena. For instance, to explore complex ECM patterning in both normal and pathological tissues, Wershof et al.~\cite{wershof2019matrix} modelled cells as diamond-shaped agents and the ECM as an orientation field on a grid with eight possible directions. Their model incorporates random motion, cell--cell and ECM guidance, collagen secretion, degradation, and rearrangement. The authors demonstrate how basic rules governing fibroblast--ECM interactions can give rise to complex tissue patterns. Sharan et al.~\cite{poonja2023dynamics} developed an off-lattice hybrid model, where cells are represented as discrete agents and the ECM as a unit vector field, to study contact guidance and cell-generated forces on the ECM, which lead to different ECM fibril alignments and tumour-associated collagen signatures. To predict neovessel guidance during angiogenesis, Labelle et al.~\cite{labelle2023spatial} developed computational models where collagen is represented as a deformable three-dimensional ellipsoidal fibril distribution, accounting for the interplay between collagen orientation, anisotropy, density, and vessel alignment. While these models provide a robust framework for incorporating diverse biophysical phenomena, their complexity often limits the systematic parameter studies that are necessary to isolate and understand specific mechanisms. 


\subsection{Aims and outline}

The primary aim of this work is to develop a computationally tractable mathematical model that focuses exclusively on the role of ECM-generated contact guidance, enabling detailed exploration of how collagen fibres influence cell migration and \textit{vice versa}, whilst at the same time incorporating biologically realistic cell--cell interactions, such as volume filling and cell--cell adhesion. We perform a comprehensive model parameter sweep to explore how collective cell invasion signatures are shaped by cell phenotype.

The minimal model we develop provides advances over existing models by integrating well-established, individual-based mathematical representations of cell--cell interactions, such as cell--cell adhesion and volume filling, with the continuous tensorial description of collagen fibres proposed by Cumming et al.~\cite{cumming2010mathematical}. We focus on the role of contact guidance in influencing cell motility, and the impact of cells in remodelling the collagen field. For simplicity, we assume that contact guidance modulates the direction of cell migration without affecting the speed, though this feature could easily be included in future iterations of the model. A systematic exploration of the model predictions across different regions of parameter space enables us to explore how cell phenotype impacts invasion dynamics.

The structure of the paper is as follows: In Section~\ref{sec:methods}, we outline the mathematical model and in Section~\ref{Subsection:Pseudocodes} we detail the computational implementation. Section~\ref{sec:results} presents our findings on how different mechanisms, separately and collectively, give rise to various cell and collagen fibre dynamics. Finally, we discuss the results and future directions in Section~\ref{sec:discussion}.


\section{Mathematical model}
\label{sec:methods}

We formulate a two-dimensional hybrid mathematical model of collective cell motility mediated by collagen fibres. Cells are modelled as discrete agents migrating \YY{on top of} collagen fibres, and the collagen fibres are modelled via a continuous fibre field. The model incorporates the following key mechanisms: random cell motility and cell--cell interactions, both modulated by cell--fibre interactions; cell proliferation; and collagen fibre secretion and degradation by cells. We introduce the representation of cells and collagen fibres in Section~\ref{sec:Representations}, and the equations determining their evolution are given in Section~\ref{sec:SystemDynamics}. Throughout we adopt a two-dimensional Cartesian coordinate system denoted by $\mathbf{x}:=\left(x, y\right)^\mathrm{T}$ \YY{with time represented by the independent variable} $t$. Key variables and parameters are summarised in Tables~\ref{tab:var} and~\ref{tab:ParamAndVar}, respectively, and a  schematic illustration of the model can be found in Figure~\ref{Fig1}.


\begin{table}[tbp]
\begin{center}
\begin{tabular}{ll}
\hline
\hline
\textbf{Variable} & \textbf{Description}\\
\hline
$N(t)$ & number of cells at time $t$\\
\hline
$\mathbf{X}^i(t)$ & position of cell $i$\\
\hline
$\mathbf{u}_{\rm ave}^i(\mathbf{X}^i, t; m)$ & average velocity of cell $i$ during the time interval $[t-m,t]$\\ 
\hline
$\boldsymbol{\xi}^i(t)$ & random stochastic force on cell $i$\\
\hline
$\mathbf{F}(\mathbf{X}^i-\mathbf{X}^j)$ & pairwise interaction force imposed on cell $i$ by cell $j$\\
\hline
$\beta\left(\mathbf{X}^i; \sigma\right)$ & number of cells within a distance of $2^{1/6}\sigma$ of cell $i$\\
\hline
$P_{\rm p}(\mathbf{X}^i; \Delta_0, \Delta t, \sigma)$ & probability of proliferation for cell $i$ per time step $\Delta t$\\
\hline
$\mathbf{\Omega}(\mathbf{x}, t)$ & collagen fibre orientation tensor \\
\hline
$\hat{\mathbf{v}}_1(\mathbf{x}, t)$ & major collagen fibre orientation \\
\hline
$\hat{\mathbf{v}}_2(\mathbf{x}, t)$ & minor collagen fibre orientation \\
\hline
$\lambda_1(\mathbf{x}, t)$ & area fraction of collagen fibres in the direction of $\hat{\mathbf{v}}_1$ \\
\hline
$\lambda_2(\mathbf{x}, t)$ & area fraction of collagen fibres in the direction of $\hat{\mathbf{v}}_2$ \\
\hline
$a\left(\mathbf{x}, t\right)$ & collagen fibre anisotropy degree\\
\hline
$\hat{\mathbf{M}}(\mathbf{X}^i, t)$ & contact guidance matrix for cell $i$ \\
\hline
$\omega\left(\mathbf{X}^i, \mathbf{x}; \sigma, \omega_0\right)$ & weight function describing cell $i$'s impact on collagen fibres\\
\hline
\hline
\end{tabular}
\end{center}
\caption{\textbf{Key variables in the model.} Hats are used to denoted unit vectors and normalised (length-preserving) tensors (see Supplementary Information Section~\ref{normalisation}).}
\label{tab:var}
\end{table}


\begin{table}[htb]
\begin{center}
\begin{tabular}{lll}
\hline
\hline
\textbf{Parameter }$\mathbf{\Theta}$ & \textbf{Description} & \textbf{Value}\\
\hline
$G$ & domain of interest & $[0, 360] \times [0, 360] $ $\mu\text{m}^2$ \\ 
&& $[0, 540] \times [0, 540] $ $\mu\text{m}^2$\\
\hline
$T$ & final time & $3600$ min, $5400$ min\\
\hline
$\Delta t$ & numerical time step & $1$ min\\
\hline
$D$ & cell diffusion coefficient & $0.3$ $\mu \text{m}^2\text{ min}^{-1}$\\
\hline
$\epsilon$ & magnitude of pairwise interactions & $0.1$ $\mu\text{m}^2 \text{ min}^{-1}$\\
\hline
$\sigma$ & cell diameter & $12$ $\mu$m~\cite{freitas1999nanomedicine}\\
\hline
$F_0$ & maximum repulsion & $2.4$ $\mu\text{m} \text{ min}^{-1}$\\
\hline
$r_{\rm max}$ & range of cell--cell interactions & $36$ $\mu$m~\cite{matsiaka2019mechanistic}\\
\hline
$\Delta_0$ & default cell cycle length & $1440$ min~\cite{seaman2015periodicity}\\
\hline
$\bar{\lambda}$ & half-maximal contact guidance & $0.4$\\
\hline
$\gamma$ & steepness of transition around $\bar{\lambda}$ & $10$\\
\hline
$m$ & length of the averaging window & $300$ min\\
\hline
$s$ & collagen fibre secretion rate & $[0, 0.5]$ $\text{min}^{-1}$\\
\hline
$d$ & collagen fibre degradation rate & $[0, 0.5]$ $\text{min}^{-1}$\\
\hline
\hline
\end{tabular}
\end{center}
\caption{\textbf{Key parameters in the model, represented by $\mathbf{\Theta}=\left(G, T, \ldots, s, d\right)^T$}. Values not obtained from the literature are selected to illustrate the results presented in Section~\ref{sec:results}.}
\label{tab:ParamAndVar}
\end{table}


\subsection{Representations for cells and collagen fibres}
\label{sec:Representations}

Cells are modelled as point particles, with the centre of cell $i$, for $i=1,\ldots,N(t)$, located at position $\mathbf{X}^i(t)\in G\subseteq \mathbb{R}^2$, where $N(t)$ is the number of cells at time $t$ and $G$ is the domain of interest. Cells interact via a pairwise interaction potential that captures both cell--cell adhesion and the fact that cells have finite volume (Section~\ref{sec:CellDynamics} and Figure~\ref{Fig1}(a),(b)).

The distribution of collagen fibres is represented by the tensor field $\mathbf{\Omega}(\mathbf{x},t)$, where $\mathbf{\Omega}$ is symmetric and positive semi-definite. Following Cumming et al.~\cite{cumming2010mathematical}, $\mathbf{\Omega}(\mathbf{x},t)$ takes the form
\begin{align}
\label{eq:GeneralRepOfFibre}
    \mathbf{\Omega}(\mathbf{x}, t) = \frac{1}{\pi}\int_{0 }^{\pi} \hat{\mathbf{u}}(\theta)\hat{\mathbf{u}}^\mathrm{T}(\theta)\rho(\theta, \mathbf{x}, t) \mathrm{d} \theta.
\end{align}
Here $\hat{\mathbf{u}}(\theta)=\left(\cos\theta, \sin\theta\right)^\mathrm{T}$ denotes a unit vector at angle $\theta$ with respect to the positive $x$-axis, and $\rho(\theta,\mathbf{x},t)\in[0, 1]$ is the area fraction occupied by collagen fibres, which we use to represent collagen fibre density, with orientation $\theta$ at position $\mathbf{x}$ and time $t$. Note that collagen fibres are bidirectional\footnote{\YY{This means that the two orientations of a fibre, separated by $180^{\circ}$, are indistinguishable.}}~\cite{dallon1999mathematical, hillen2006m5} so that $\theta \in [0, \pi)$. To enable computational tractability, we follow~\cite{cumming2010mathematical} and represent $\mathbf{\Omega}(\mathbf{x}, t)$ in a diagonalised form: 
\begin{align}
\label{eq:Omega}
    \mathbf{\Omega}(\mathbf{x}, t) = \lambda_1 \hat{\mathbf{v}}_1\hat{\mathbf{v}}_1^\mathrm{T} + \lambda_2 \hat{\mathbf{v}}_2\hat{\mathbf{v}}_2^\mathrm{T}, 
\end{align}
where $\hat{\mathbf{v}}_{1, 2}(\mathbf{x}, t)$ are the orthonormal eigenvectors of $\mathbf{\Omega}$ which capture the major and minor collagen fibre orientations, respectively, and $\lambda_{1}(\mathbf{x}, t)\geq \lambda_{2}(\mathbf{x}, t)\in [0, 1]$ are the associated eigenvalues which represent the area fractions of collagen fibres in directions $\hat{\mathbf{v}}_1$ and $\hat{\mathbf{v}}_2$, respectively. Note that \YY{for the total area fraction, which represents collagen fibre density, we have}
\begin{align}
     0\leq\lambda_1\left(\mathbf{x}, t\right) + \lambda_2\left(\mathbf{x}, t\right) = \frac{1}{\pi}\int_{0}^{\pi}\rho\left(\theta, \mathbf{x}, t\right) \text{d}\theta\leq1.
\end{align}
We define the anisotropy degree as $a:=1-\lambda_2/\lambda_1\in [0,1]$, where $a=1$ corresponds to perfect collagen fibre alignment along $\hat{\mathbf{v}}_{1}$, and $a=0$ to a perfectly isotropic collagen fibre distribution.


\subsection{System dynamics}
\label{sec:SystemDynamics}

In describing cell dynamics we follow standard modelling assumptions and neglect inertial effects~\cite{martinson2023dynamic, marchello2024non, bruckner2024learning, vargas2020modeling, cai2016modeling}. We describe cell and collagen fibre dynamics via the system of differential equations of the form
\begin{subequations}
\label{eq:GeneralDynamics}
\begin{align}
\frac{\mathrm{d}\mathbf{X}^i}{\mathrm{d}t} &= \mathbf{f}\left(\mathbf{X}^1, \mathbf{X}^2, \ldots, \mathbf{X}^{N(t)}, \mathbf{\Omega};\mathbf{\Theta}\right)\text{ and 
 } \, \mathbf{X}^i(t=0)=\mathbf{X}^i_0 \, \text{ for } i = 1, \ldots,  N(t),\label{eq:cell dynamics}\\
\frac{\partial\mathbf{\Omega}}{\partial t} &= \mathbf{g}\left(\mathbf{X}^1, \mathbf{X}^2, \ldots, \mathbf{X}^{N(t)}, \mathbf{\Omega};\mathbf{\Theta}\right)\text{ and } \, \mathbf{\Omega}(\mathbf{x}, t=0) = \mathbf{\Omega}_0(\mathbf{x}).
\label{eq:FibreFynamics}
\end{align}
\end{subequations}
The model is defined for $\mathbf{x}, \mathbf{X}^i \in G$, and $\mathbf{\Theta}$ is a vector of model parameters (see Table~\ref{tab:ParamAndVar}). The function $\mathbf{f}$ captures cell dynamics due to random motion and cell--cell interactions, both influenced by contact guidance from the surrounding collagen fibres. The function $\mathbf{g}$ captures the remodelling of collagen fibres via degradation and secretion by surrounding cells. In Sections~\ref{sec:CellDynamics}--\ref{Subsection:Fibre dynamics} we are provide specific functional forms for $\mathbf{f}$ and $\mathbf{g}$.


\subsubsection{Cell dynamics}
\label{sec:CellDynamics}
 
The equation of motion for each cell $i=1,\ldots,N(t)$ is given by Equation (\ref{eq:cell dynamics}) with 
\begin{equation}
\label{cellDynamics}
    \mathbf{f}\left(\mathbf{X}^1, \mathbf{X}^2, \ldots, \mathbf{X}^{N(t)}, \mathbf{\Omega};\mathbf{\Theta}\right)=\hat{\mathbf{M}}\left(\mathbf{X}^i,t\right)\left[\boldsymbol{\xi}^i+ \sum_{j=1,j\neq{i}}^{N(t)}\mathbf{F}\left(\mathbf{X}^i-\mathbf{X}^j\right)\right],
\end{equation}
where the $2\times 1$ vector $\boldsymbol{\xi}^i$ models random motility, the $2\times 1$ vector $\mathbf{F}\left(\mathbf{X}^i-\mathbf{X}^j\right)$ captures the pairwise interactions between cell $i$ and cell $j$, and the $2\times 2$ normalised (length-preserving) matrix $\hat{\mathbf{M}}$ models the modulation of cell motility due to contact guidance from the underlying collagen fibres.


\paragraph{Random motility and cell--cell interactions.}
We assume the random motion of cell $i$ can be captured by a stochastic force $\boldsymbol{\xi}^i$~\cite{matsiaka2019mechanistic,nava2017extracting,porta2019statistical,berg1993random}. We implement $\boldsymbol{\xi}^i$ by sampling from a Gaussian distribution with zero mean and variance $2D/\Delta t$, where $D>0$ is the macroscopic cell diffusion coefficient and $\Delta t>0$ is the numerical time step.

The pairwise interaction force experienced by cell $i$ as a result of interactions with cell $j$ is of the form\begin{equation}
\label{eq:c-cKernel}
    \mathbf{F}\left(\mathbf{X}^i-\mathbf{X}^j\right) = F\left(|\mathbf{X}^i-\mathbf{X}^j|\right) \frac{\mathbf{X}^i- \mathbf{X}^j}{|\mathbf{X}^i - \mathbf{X}^j|},
\end{equation}
where $F\left(|\mathbf{X}^i-\mathbf{X}^j|\right)$ is the magnitude of the force, which acts in the direction $\mathbf{X}^i- \mathbf{X}^j$. We use the Lennard-Jones potential~\cite{wang2020lennard,lennard1924determination} so that
\begin{align}
    \begin{split}
    F = \begin{cases}
        \displaystyle \min\left\{24\epsilon \sigma^6\left[\frac{2\sigma^6}{|\mathbf{X}^i - \mathbf{X}^j|^{13}} - \frac{1}{|\mathbf{X}^i - \mathbf{X}^j|^7}\right], \; F_0\right\}, \quad &  0 < |\mathbf{X}^i - \mathbf{X}^j| \leq r_{\rm max},\\
        \displaystyle 0, \quad &  |\mathbf{X}^i - \mathbf{X}^j| > r_{\rm max}.
    \end{cases}
    \end{split}\label{eq:cell--cellInteractionForce} 
\end{align}
Here $\sigma>0$ is a measure of the cell diameter, and $\epsilon>0$ regulates the magnitude of the pairwise interaction forces. The additional term $F_0>0$ prevents $F$ from approaching infinity as $|\mathbf{X}^i - \mathbf{X}^j|\to0$ and, in line with previous works, we take $r_{\rm max}=3\sigma>0$~\cite{matsiaka2019mechanistic}.


\paragraph{Contact guidance.}

To model contact guidance, we take
\begin{equation}
    \label{eq:M_rm}
    \mathbf{M}(\mathbf{X}^i, t)= \Lambda\left(\lambda_1+\lambda_2\right)\hat{\mathbf{\Omega}}\left(\mathbf{X}^i, t\right) + \left(1-\Lambda\left(\lambda_1+\lambda_2\right)\right) \mathbf{I},
\end{equation}
where $\mathbf{I}$ is the $2 \times 2$ identity matrix and $\hat{\mathbf{\Omega}}$ is the normalised version of $\mathbf{\Omega}$, as defined in Supplementary Information Equation~\eqref{omega normal}. We then apply $\hat{\mathbf{M}}$ (the normalised version of $\mathbf{M}$, defined in the same way---see Supplementary Information Section~\ref{normalisation}) to the right-hand side of Equation~\eqref{cellDynamics} to ensure that contact guidance impacts only the direction of motion, and not cell speed.

The strength of contact guidance is weighted by a function $\Lambda$ which depends on the total area fraction of collagen fibres, $\lambda_1+\lambda_2$, and is taken to be
\begin{equation}
    \label{eq:Lambda}
    \Lambda(\lambda_1+\lambda_2):=\frac{h\left(\lambda_1+\lambda_2\right)-h(0)}{h(1) -h(0)},\text{ where } h(x):=\frac{1}{2} \left(\tanh\left(\gamma \left(x-\bar{\lambda}\right)\right)+1\right).
\end{equation}
The parameter $\bar{\lambda}\in[0,1]$ represents the critical value of $\lambda_1+\lambda_2$ at which the strength of contact guidance is half maximal, and $\gamma > 0$ controls the steepness of the transition around $\bar{\lambda}$.

\begin{figure}[htbp]
\centering
    \includegraphics[width=0.9\linewidth]{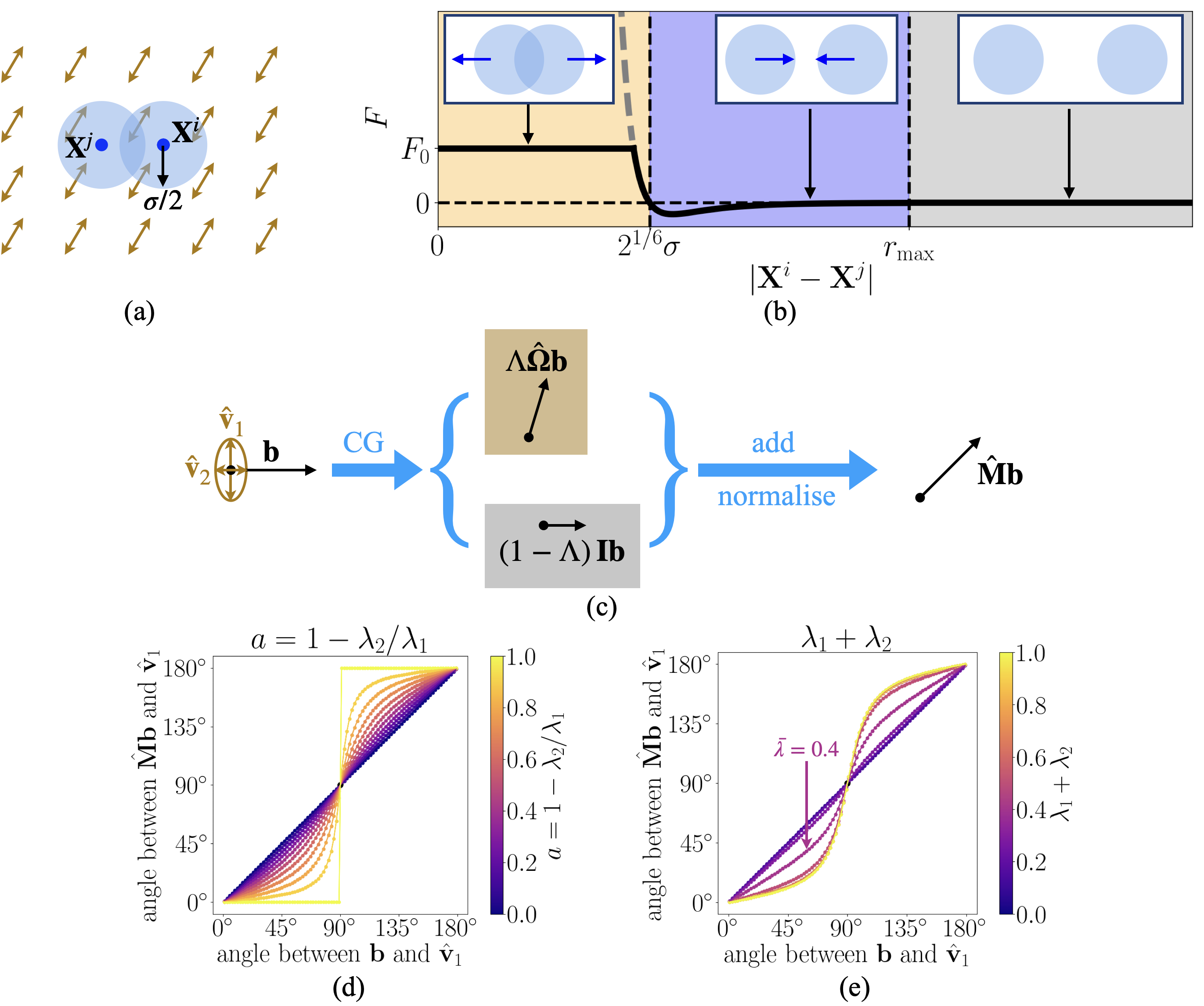}
    \caption{\textbf{Schematic illustration of the model.} (a) Cells $i$ and $j$ are depicted as blue circular discs with radius $\sigma/2$ centred at $\mathbf{X}^i$ and $\mathbf{X}^j$, respectively. They sit on top of collagen fibres, where major fibre orientation $\hat{\mathbf{v}}_1$ and the associated area fraction $\lambda_1$ are visualised using brown arrows. (b) The magnitude of the pairwise interaction force is given by the function $F(|\mathbf{X}^i- \mathbf{X}^j|)$ defined in Equation~\eqref{eq:cell--cellInteractionForce}. Orange region: short-range repulsion due to the effects of volume filling; blue region: mid-range adhesion; grey region: no cell--cell interactions. The dashed grey line represents the force arising from the Lennard-Jones potential, and $F_0>0$ sets the maximum repulsive force used in the model. (c) Schematic diagram of how the contact guidance matrix $\hat{\mathbf{M}}$ reorients vector $\mathbf{b}$ given vertical $\hat{\mathbf{v}}_1$ and horizontal  $\hat{\mathbf{v}}_2$. (d),(e) Illustration of the mechanisms of contact guidance indicating how $\hat{\mathbf{M}}$, defined in Equation~\eqref{eq:M_rm}, reorients a vector $\mathbf{b}$, given different anisotropy degrees, $a$ (d), and total area fractions, $\lambda_1+\lambda_2$ (e). In (d), $\lambda_1+\lambda_2=1.0$ and in (e), $a=0.8$. \YY{In (e), the arrow indicates the strength of contact guidance at the critical total fibre area fraction $\bar{\lambda}=0.4$.}} 
\label{Fig1}
\end{figure}

The choice of $\Lambda$ ensures key properties are satisfied:~when the space is saturated with collagen fibres ($\lambda_1+\lambda_2=1$), $\Lambda = 1$ and $\hat{\mathbf{M}}=\hat{\mathbf{\Omega}}$, and the impact of contact guidance is maximal; whereas in the absence of collagen fibres, $\Lambda = 0$ and $\hat{\mathbf{M}}=\mathbf{I}$ so that the cells experience no contact guidance. Figure~\ref{Fig1}(c) visualises how $\hat{\mathbf{M}}$ reorients any $2 \times 1$ vector $\mathbf{b}$, and Figure~\ref{Fig1}(d)--(e) compares the angle between $\mathbf{b}$ and $\hat{\mathbf{v}}_1$ (the major orientation of collagen fibres) both with and without contact guidance. As shown in Figure~\ref{Fig1}(d), $\hat{\mathbf{M}}$ automatically encodes the anisotropy information $a$. As $a$ approaches zero (isotropic collagen fibres), $\hat{\mathbf{M}}$ tends towards the identity matrix, $\mathbf{I}$, and there is no contact guidance. Moreover, Figure~\ref{Fig1}(e) indicates that there is a switch between strong and weak contact guidance around $\lambda_1+\lambda_2=\bar{\lambda}$, as expected. \YY{Note that Figure~\ref{Fig1}(d)–(e) are symmetric around $90^\circ$ because collagen fibres are bidirectional. Figure~\ref{Fig1} also illustrates the advantage of representing the collagen fibres as a tensorial field: the matrix-vector multiplication ($\hat{\mathbf{M}}\mathbf{b}$) directly reorients the cell’s migratory direction without the need for additional assumptions to select between the two directions (see Supplementary Information Section~\ref{normalisation}), due to the bidirectionality of collagen fibres.}


\subsubsection{Cell proliferation}
\label{sec:proliferation}

Following previous work~\cite{tarle2015modeling}, we implement a model of cell density-dependent cell proliferation in which the probability that a cell divides in the time step $[t,t+\Delta{t})$ is given by
\begin{align}
\label{eq:cell_prolif}
    P_{\rm p}(\mathbf{X}^i(t);\Delta_0,\Delta t,\sigma)= 
    \begin{cases}
        \displaystyle \frac{\Delta t}{\Delta_0}\left(1-\frac{\beta\left(\mathbf{X}^i(t); \sigma\right)}{6}\right), &\quad \beta \leq 6,\\
         0, &\quad \text{otherwise,}
    \end{cases}
\end{align}
where $\Delta_0>0$ is the mean cell cycle length in the absence of crowding, $\beta\left(\mathbf{X}^i(t), \sigma\right)$ is the number of cells in the region in which the pairwise interaction force is repulsive, and six is assumed to be the maximum packing number~\cite{hales2017formal}. 


\subsubsection{Collagen fibre dynamics}
\label{Subsection:Fibre dynamics}

We assume that cells evolve the collagen fibres through degradation and secretion. The \YY{governing} equation for collagen fibres is given by Equation (\ref{eq:FibreFynamics}), where $\mathbf{g}$ is as follows:
\begin{equation}
    \label{eq:FibreDynamicsEq}
    \mathbf{g}\left(\mathbf{X}^1, \mathbf{X}^2, \ldots, \mathbf{X}^{N(t)}, \mathbf{\Omega};\mathbf{\Theta}\right) = \sum_{i = 1}^{N(t)}\omega\left(\mathbf{X}^i, \mathbf{x};\sigma\right) \left[s\left(1 - \lambda_1 - \lambda_2\right)\hat{\mathbf{u}}_{\rm ave}^i \left(\hat{\mathbf{u}}_{\rm ave}^i\right)^\mathrm{T} - d\,\mathbf{\Omega}\right],
\end{equation}
where $s>0$ and $d>0$ are the collagen secretion and degradation rates per cell, respectively. The term $\left(1 - \lambda_1 - \lambda_2\right)$ ensures no fibre secretion when the space is fully occupied by collagen fibres. The secretion of collagen fibres by cell $i$ is in the direction of cell $i$'s average velocity, $\mathbf{u}_{\rm ave}^i$, over $[t-m,t]$, where\footnote{We set $\mathbf{X}^i(t-m)=\mathbf{X}^i(0)$ if $t<m$, where $\mathbf{X}^i(0)$ is the initial position for cell $i = 1$.}
\begin{align}
\label{ave tau previous vel}
    \mathbf{u}_{\rm ave}^i(\mathbf{X}^i, t; m)=\frac{1}{m}\int_{t-m}^{t}\frac{\mathrm{d}\mathbf{X}^i}{\mathrm{d}t'}\mathrm{d}t'=\frac{1}{m}\left(\mathbf{X}^i(t)-\mathbf{X}^i(t-m)\right).
\end{align}
In writing Equation~\eqref{eq:FibreDynamicsEq}, we assume a cell impacts the distribution of collagen fibres within a distance $\sigma/2$ from its centre, recalling $\sigma$ captures the cell diameter. To reflect this, we take
\begin{equation}
    \label{eq:cell-fibreWeightFunction}
    \omega\left(\mathbf{X}^i, \mathbf{x},\sigma\right) = 
    \begin{cases}
    1-\frac{|\mathbf{X}^i-\mathbf{x}|}{\sigma/2}, & \quad |\mathbf{X}^i-\mathbf{x}|\leq \sigma/2,\\
    0, & \quad |\mathbf{X}^i-\mathbf{x}|> \sigma/2.
    \end{cases}
\end{equation}


\section{Computational implementation of the model}\label{Subsection:Pseudocodes}

In this paper, we use two distinct two-dimensional rectangular domains, $G$, and final times, $T$, to illustrate the results presented in later Section~\ref{sec:results}:
\begin{enumerate}[label={}, leftmargin=\parindent]
\item \textbf{Setup 1.}  $G = [0, 540] \times [0, 540] \, \mu\mathrm{m}^2$ and $T = 96$ hours, used to showcase the main features of the model \YY{in detail}.\label{setup1}
\item \textbf{Setup 2.}  $G = [0, 360] \times [0, 360] \, \mu\mathrm{m}^2$ and $T = 60$ hours \YY{(smaller domain size and longer simulation time for computational efficiency)}, used to investigate the effects of different collagen fibre initial conditions and cell properties on dynamics.\label{setup2}
\end{enumerate}
We impose periodic boundary conditions for the cells, and the domain $G$ is discretised as $G_\delta$ with a grid size of $\delta = 2 \, \mu\mathrm{m}$, and the numerical time step is set to $\Delta{t}=1$ minute. The initial conditions are listed in Supplementary Information Section \ref{Sec:ICS}. The collagen fibre field, $\mathbf{\Omega}$, is discretised on $G_\delta$ as $\mathbf{\Omega}_\delta$. At each time step we:
\begin{enumerate}
    \item \sloppy calculate cell $i$'s random motility force $\boldsymbol{\xi}^i$ and cell--cell interaction forces $\sum_{j = 1, j \neq i}^{N(t)}\mathbf{F}\left(\mathbf{X}^i-\mathbf{X}^j\right)$ (see Section~\ref{sec:CellDynamics});
    \item extract $\mathbf{\Omega}_\delta$ at cell $i$'s centre using linear interpolation to obtain the matrix $\hat{\mathbf{M}}$ for contact guidance (see Equation~\eqref{eq:M_rm}).
\end{enumerate}
We then update all cell positions according to Equation~\eqref{cellDynamics}, and update the discretised collagen fibre field $\mathbf{\Omega}_\delta$ according to Equations (\ref{eq:FibreDynamicsEq})--(\ref{ave tau previous vel}).
Finally, we iterate over the $N(t)$ cells to let each cell proliferate with probability $P_{\rm p}$ defined in Equation~\eqref{eq:cell_prolif}. The daughter cell is placed a distance $\sigma/2$ away at an angle sampled from the uniform distribution $U[0, 2\pi]$.

Algorithm~\ref{pseudocode} provides the pseudocode for the numerical simulation of the model. The differential equations are solved using the forward Euler method~\cite{hairer1987solving} with time step $\Delta t$, and the integrals are approximated using the midpoint rule~\cite{zorn2002calculus}. Initial conditions for each simulation in Section~\ref{sec:results} are stated and visualised in Supplementary Information (Section~\ref{Sec:ICS}). Python codes to simulate the model are available at \url{https://github.com/YuanYIN99/ECMcell_ContactGuidance_HybridModel.git}

\bigskip


\begin{algorithm}[h]
  \caption{Collective cell migration in a fibrous environment.}
  
  \begin{algorithmic}[1]
    \medskip
    \State \textbf{Input}: \\
    Initialisation: discretised rectangular domain $G_\delta$; current time $t=0$, time step $\Delta t$, and final time $T$; $N(t)$ cell locations $\mathbf{X}^1(t), \mathbf{X}^2(t), \ldots, \mathbf{X}^{N(t)}(t)$; discretised collagen fibre field $\mathbf{\Omega}_{\delta}$; parameters $\mathbf{\Theta}$ listed in Table~\ref{tab:ParamAndVar}.
    
    \State

    \While{$t + \Delta t \leq T$} \\

    \For{cell $i$ in $N(t)$ cells, sequentially}
      \If{cell $i$ at position $\mathbf{X}^i(t)$ has fewer than six repulsive neighbours}
        \State Let cell $i$ proliferate with probability $P_{\rm p}$ given by Equation~\eqref{eq:cell_prolif}.
        \EndIf
      \EndFor\\
    
      \For{all cells \( i =1, 2, \ldots, N(t) \), in parallel}
      \State Sample $\boldsymbol{\xi}^i$ 
      (see Section~\ref{sec:CellDynamics}).

      \State \parbox[t]{\dimexpr\linewidth-\algorithmicindent}{Loop through cells $j \neq i$ and calculate $\sum_{j = 1, j \neq i}^{N(t)}\mathbf{F}\left(\mathbf{X}^i-\mathbf{X}^j\right)$ (see Section~\ref{sec:CellDynamics}).\strut}

       \State \parbox[t]{\dimexpr\linewidth-\algorithmicindent}{Extract $\mathbf{\Omega}_\delta$ at $\mathbf{X}^i$ and calculate the contact guidance matrix $\hat{\mathbf{M}}$ by Equation~\eqref{eq:M_rm}.\strut}

       \State \parbox[t]{\dimexpr\linewidth-\algorithmicindent}{Migrate cell $i$ based on Equation~\eqref{cellDynamics} to obtain $\mathbf{X}^i( t+\Delta t)$. Record its average velocity according to Equation~\eqref{ave tau previous vel}.\strut}
      
      \EndFor\\

      \For{cell $i$ in $N(t)$ cells, sequentially}

        \State \parbox[t]{\dimexpr\linewidth-\algorithmicindent}{Find the grid points that are impacted by cell $i$ at $\mathbf{X}^i(t)$ through secretion and degradation. Update the collagen fibre distribution dynamics on those grid points based on Equation~\eqref{eq:FibreDynamicsEq} to obtain $\mathbf{\Omega}_{\delta}(\mathbf{x}, t+\Delta t)$.\strut}
        
      \EndFor\\

      \State Step in time: $t = t + \Delta t$.
    \EndWhile\\
    \State \textbf{Return} $\mathbf{X}^i$ and $\mathbf{\Omega}_{\delta}$ at times $t=0, \Delta t, 2\Delta t, \ldots, T$.
  \end{algorithmic}
  \medskip
  \label{pseudocode}
\end{algorithm}


\section{Results}
\label{sec:results}

We now demonstrate how differences in cell phenotypes impact patterns of collective cell invasion, underpinned by the interactions between cells and collagen fibres. We first, in Section~\ref{sec:4.1}, explore the collagen fibre distributions that arise \textit{de novo} from a migratory cell phenotype, before moving to understand how different patterns of collagen fibres drive collective cell invasion in a phenotype that does not re-model the underlying fibre bed in Section~\ref{sec:4.2}. We then, in Section~\ref{sec:4.3}, simulate the dynamics of \YY{degrading and secreting} phenotypes in turn, to tease apart the subtleties of the two-way coupling between cells and collagen fibres. In Section~\ref{sec:4.4}, we conduct a systematic parameter sweep to reveal how a cell phenotype able to fully re-model collagen via both secretion of new fibres and degradation of old fibres can undergo collective invasion. Finally, in Section~\ref{sec:4.5} we showcase the wide range of possible invasion behaviours as the relative contributions of different processes are varied. \YY{All cell phenotypes and their corresponding dynamics explored are listed in Table \ref{tab:phenotypes}.}

\begin{table}[tbh]
    \centering
    \begin{tabular}{|l|l|}
    \hline
    \hline
    \textbf{Cell phenotypes} & \textbf{Section} \\ \hline
    Non-collagen modulating phenotype & Section 4.2 \\ \hline
    Collagen degrading phenotype & \multirow{2}{*}{Section 4.3} \\ \cline{1-1}
    Collagen secreting phenotype &  \\ \hline
    Collagen-modulating phenotype & Section 4.4 \\ \hline
    Cell-cell interaction migratory phenotype & \multicolumn{1}{c|}{\multirow{4}{*}{Section 4.5}} \\ \cline{1-1}
    Random motility migratory phenotype & \multicolumn{1}{c|}{} \\ \cline{1-1}
    Contact guidance (in)sensitive phenotype(s) & \multicolumn{1}{c|}{} \\ \cline{1-1}
    Short (\& long) memory phenotype(s) & \multicolumn{1}{c|}{}
    \\ 
    \hline
    \hline
    \end{tabular}
    \caption{\textbf{\YY{Cell phenotypes and their corresponding dynamics explored in Section 4.}}}
    \label{tab:phenotypes}
    \end{table}


\subsection{Secreted collagen fibre dynamics reflect cell migration patterns}
\label{sec:4.1}

\begin{figure}[htbp]
\centering
\includegraphics[width=1.0\textwidth]{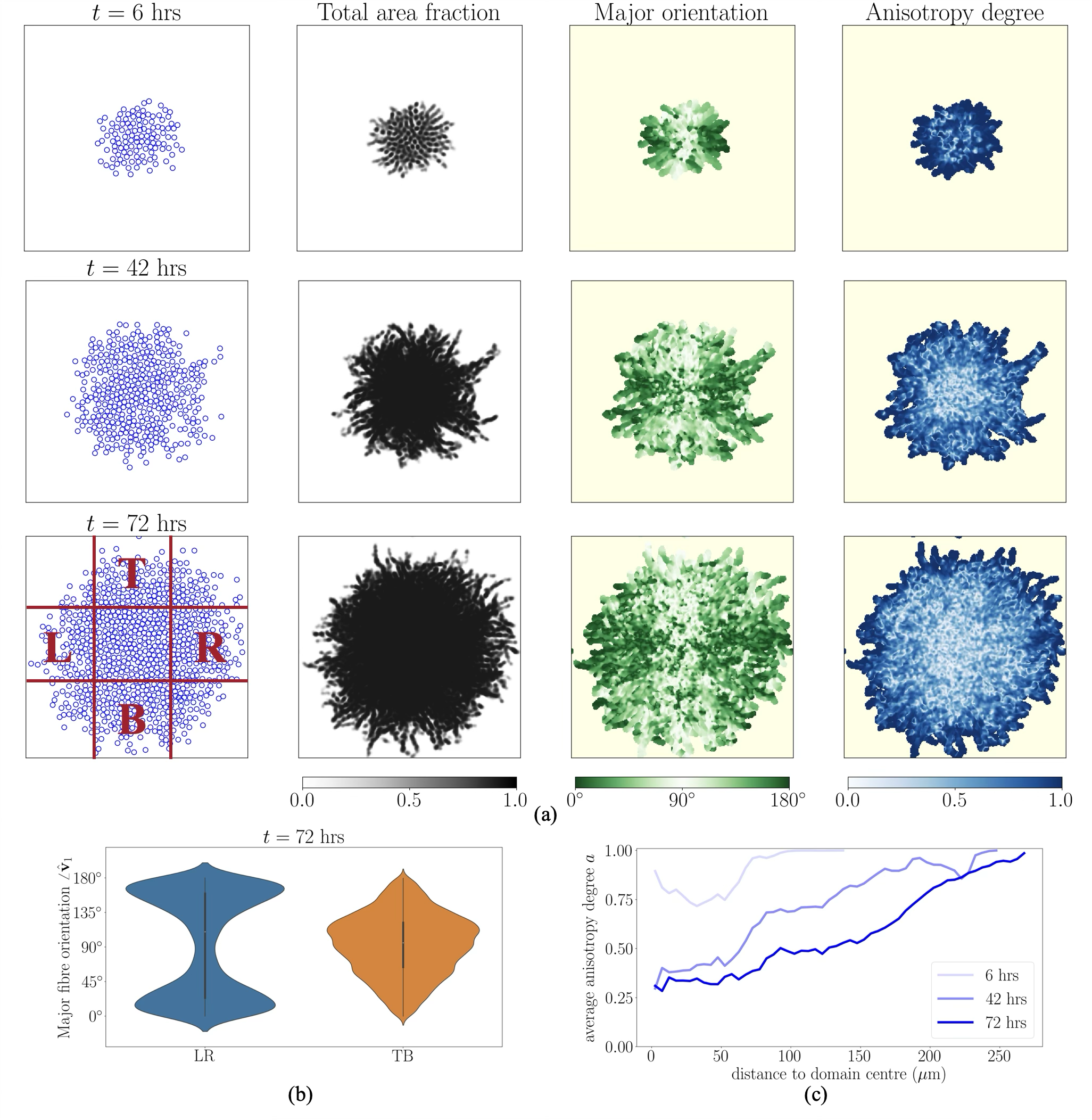}
\caption{\textbf{Secreted collagen fibre dynamics reflect cell migration patterns.} (a) Cell and collagen fibre distributions at $6$, $42$, and $72$ hours. Cells are visualised as blue circles with radius $\sigma/2$. The pale yellow backgrounds denote regions devoid of collagen fibres. (b) Distribution of the angle between $\hat{\mathbf{v}}_1$ and the positive $x$-axis in the left and right regions (LR) compared with the top and bottom regions (TB) (the regions `L', `R’, `T', and `B’ are indicated in the bottom-left of (a)). \YY{The violin plot shows the kernel density estimate of this distribution, with a miniature box plot overlaid: the thick black bar represents the interquartile range (IQR, Q1--Q3), the white dot indicates the median, and the thin black lines (`whiskers') extend to data points within $1.5\times$IQR from Q1 and Q3.} (c) Average anisotropy degree within concentric rings from the domain centre. Simulation details: Setup1 in Section~\ref{Subsection:Pseudocodes}. Initially, $100$ cells are arranged in a densely packed circular disc at the centre of the domain, forming a confluent arrangement, with no collagen fibres present ($\lambda_1+\lambda_2 = 0$). Collagen fibre secretion and degradation rates are $0.025\text{ min}^{-1}$ and $0.0025 \text{ min}^{-1}$, respectively.} 
\label{Fig2}
\end{figure}

In Figure~\ref{Fig2}, we showcase how collective cell invasion leads to patterns in collagen fibre dynamics. We take as initial conditions $100$ cells clustered in the centre of the domain, with no collagen fibres present. As the cells migrate outwards and proliferate, they lay down collagen fibres that are oriented in the direction of outward expansion. Figure~\ref{Fig2}(a) shows snapshots of the cell positions, the total area fraction of collagen fibres ($\lambda_1 + \lambda_2$), the major orientation ($\hat{\mathbf{v}}_1$), and the anisotropy degree ($a$), shown in grey, green, and blue shading, respectively. Cells at the leading edge exhibit directed motion due to population pressure, resulting in the secretion of highly aligned collagen fibres (Figure~\ref{Fig2}(b)). Cells behind the front are limited in their motility due to the lack of free space leading to a more isotropic collagen fibre distribution (Figure~\ref{Fig2}(c)). \YY{We note that the dynamics of cells and collagen fibres are qualitatively insensitive to the functional form of the collagen-modulating kernel, $\omega$ (see Figures~\ref{Fig:heaviside_w}--\ref{fig:nonlocal_w}).}


\subsection{Non-ECM-modulating phenotypes display restricted invasion into dense collagen environments}
\label{sec:4.2}

\begin{figure}[htbp]
\centering
\includegraphics[width=1.0\textwidth]{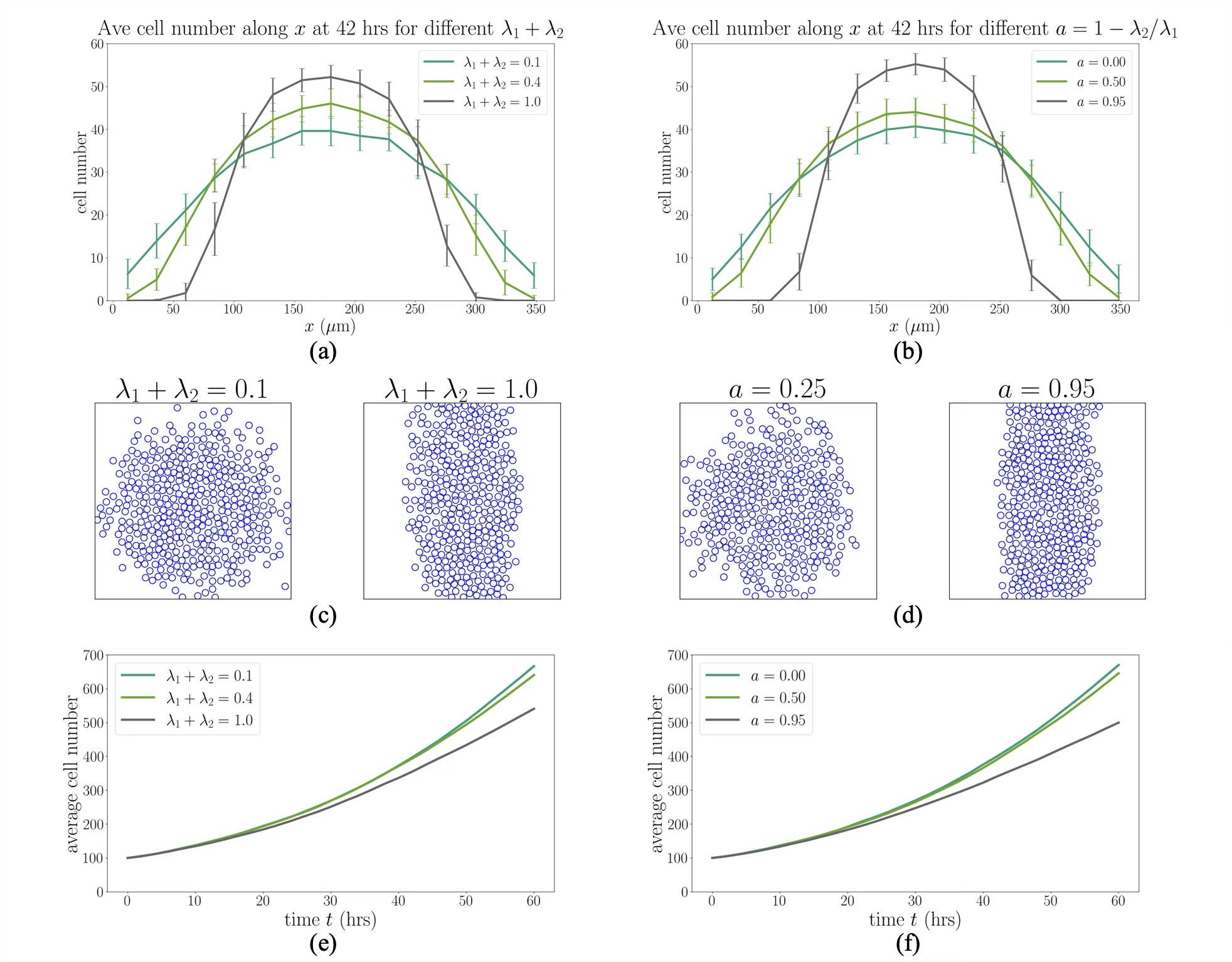}
\caption{\textbf{Non-ECM-modulating phenotypes display restricted invasion into dense collagen environments.} (a),(b) One-dimensional cell density profiles obtained by counting the number of cells along the $x$-direction and averaging over the $y$-direction at $42$ hours. Cell density is shown for different total area fractions ($\lambda_1 + \lambda_2$) and anisotropy degrees ($a = 1 - \lambda_2 / \lambda_1$) of collagen fibres. Error bars represent the standard deviation over $40$ repetitions. (c),(d) Cell distributions at $42$ hours. (e),(f) Average cell number over time. For (a), (c), and (e), $a=0.9$ while for (b), (d), and (f), $\lambda_1+\lambda_2 = 0.8$. In all figures, the major fibre orientation, $\hat{\mathbf{v}}_1$, is vertical while the minor orientation, $\hat{\mathbf{v}}_2$, is horizontal. Simulation details: Setup2 in Section~\ref{Subsection:Pseudocodes}. Initially, $100$ cells are arranged in a densely packed circular disc at the centre of the domain, forming a confluent arrangement. Collagen fibre secretion and degradation rates are both set to zero.}
\label{Fig3}
\end{figure}

We now explore collective cell invasion in a non-ECM-modulating phenotype where cells respond to contact guidance cues from the underlying collagen fibres without altering them ($s=d=0$ $\text{min}^{-1}$). Here we again simulate using an initial cluster of 100 cells, but now on a spatially-uniform fibre bed with major fibre orientation, $\hat{\mathbf{v}}_1$, vertical and minor fibre orientation, $\hat{\mathbf{v}}_2$, horizontal. On the left-hand side of Figure \ref{Fig3}, we show results where the collagen fibres are primarily aligned vertically and the total area fraction of collagen varies, whilst on the right-hand side, the anisotropy is varied whilst the total area fraction of collagen is held fixed (again, with primarily vertically aligned collagen fibres). 

As expected, Figure~\ref{Fig3} shows that both density and anisotropy of the collagen fibre field impact cell invasion patterns. On the left-hand side, where the anisotropy is fixed at $a=0.9$, we see that a low area fraction of collagen fibres does not restrain collective invasion, whereas a high area fraction restricts cell migration horizontally. On the right-hand side of Figure \ref{Fig3}, where the total area fraction of collagen fibres is fixed at $\lambda_1+\lambda_2=0.8$, the cell migration patterns reflect the underlying anisotropy of the fibre field (more anisotropy confines cells horizontally). 

Another important prediction of the model is that the initial collagen fibre distribution indirectly influences cell proliferation. This occurs because the cell clusters are more compact when the collagen fibres are more dense and more aligned (Figure~\ref{Fig3}(c),(d), and Figure~\ref{fig:CG-ONLY_supp} in the Supplementary Information). This effect is not apparent until approximately $24$ hours \YY{from the start of the simulation} (Figure~\ref{Fig3}(e),(f)), which corresponds to the average cell cycle length.


\subsection{Collagen-modulating phenotypes display different invasion patterns}
\label{sec:4.3}

\begin{figure}[htbp]
\centering
\includegraphics[width=1.0\textwidth]{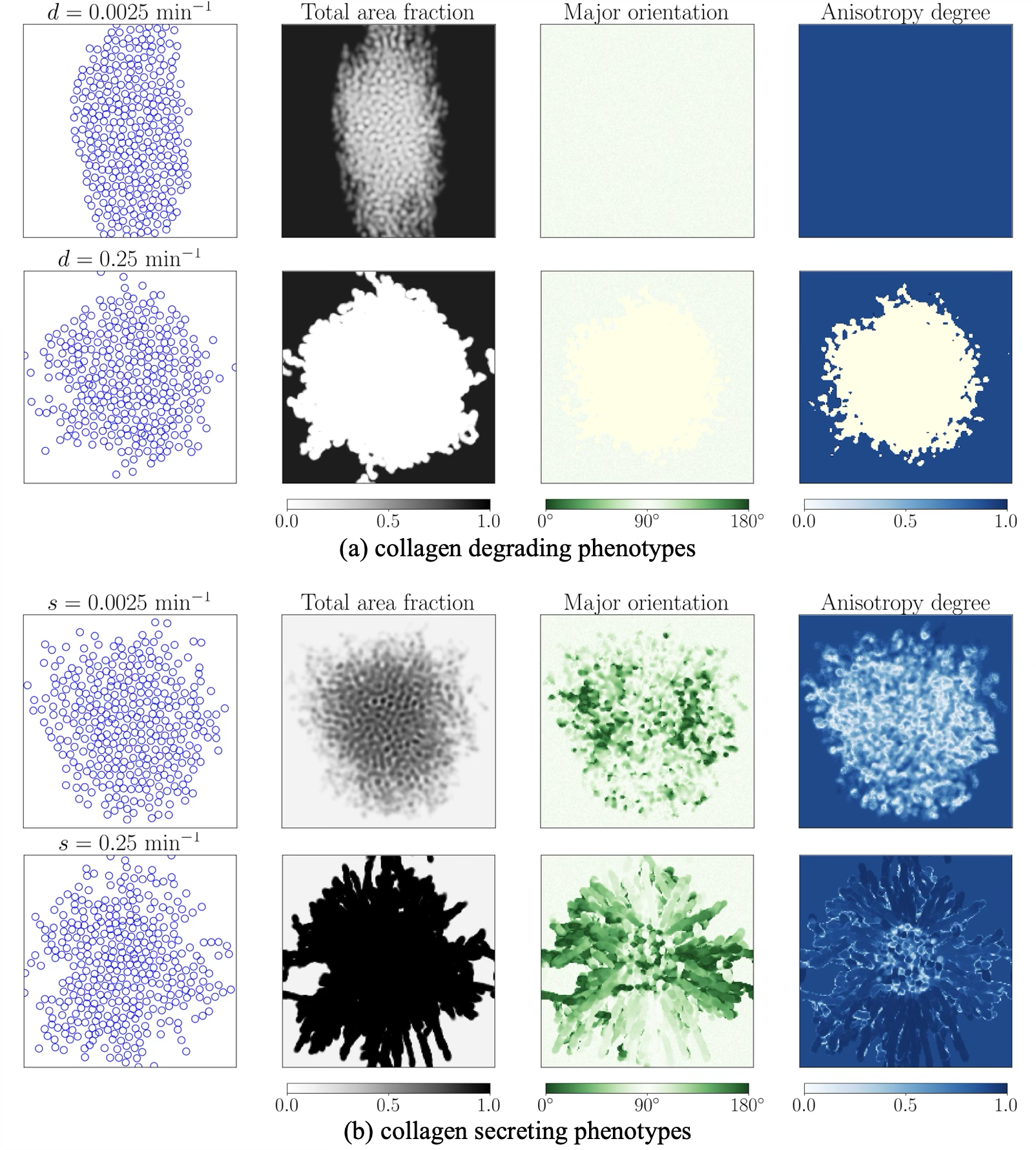} \\
\vspace{0.8cm}
\caption{\textbf{Dynamics of collagen degrading and secreting phenotypes.} (a) collagen degrading phenotypes display enhanced invasion. Here, the initial conditions have $\lambda_1+\lambda_2=0.9$, $a=1-\lambda_2/\lambda_1=0.9$, vertical $\hat{\mathbf{v}}_1$ and horizontal $\hat{\mathbf{v}}_2$. (b): collagen-secreting phenotypes promote anisotropic cell invasion and collagen patterns. Here the initial conditions have $\lambda_1+\lambda_2=0.1$, $a=1-\lambda_2/\lambda_1=0.9$, vertical $\hat{\mathbf{v}}_1$ and horizontal $\hat{\mathbf{v}}_2$. Simulation setup: Setup2 in Section~\ref{Subsection:Pseudocodes}. The initial conditions are visualised in Figure~\ref{fig:ICs} in Supplementary Information (Section~\ref{Sec:ICS}), with $100$ cells, and results are displayed at $t=36$ hours. The pale yellow shading denotes a region devoid of collagen fibres.} 
\label{Fig4}
\end{figure}

To examine the dynamics of collagen fibre degrading phenotype we set the secretion rate to zero, $s=0$. Figure~\ref{Fig4}(a) shows that faster collagen fibre degradation reduces contact guidance, with cells displaying invasion patterns similar to when no fibres are present (compare with Figure~\ref{Fig2}). We observe that a higher degradation rate promotes proliferation by facilitating dispersion of the cells (see Figure~\ref{fig:fig4_SI}(a),(b) in Supplementary Information (Section~\ref{sec: fig4_SI})).


We now examine the dynamics of a collagen-secreting phenotype, setting the degradation rate to be zero, $d=0.0$, and the initial total area fraction of collagen to be low so that there is substantial area available into which cells can secrete new collagen. Here, the remodelling process is more intricate, and displays spatial variations across the cluster (see Figure~\ref{Fig4}(b)). For cells in the top and bottom regions (as defined in Figure~\ref{Fig2}), migration is primarily aligned with the vertically oriented major fibre direction. In these regions, anisotropy increases over time as newly secreted fibres reinforce this vertical alignment. For cells in the left and right regions (as defined in Figure~\ref{Fig2}), migration is primarily driven by population pressure arising from cell--cell interactions, and is aligned with the horizontally oriented minor fibre direction. As the cells secrete new collagen, which is primarily horizontally aligned, the fibre anisotropy initially decreases over time as the fibre distribution becomes more uniform (i.e., $\lambda_1 \approx \lambda_2$). It then increases as newly secreted fibres are continually laid down in the horizontal alignment, so that the major fibre direction becomes horizontal. In the central region, the anisotropy decreases over time as cells move predominantly randomly. \YY{Therefore, as shown in Figure~\ref{Fig4}(b), the high collagen-secreting cell phenotype deposits denser collagen fibres in a more anisotropically aligned structure compared with the low-secreting phenotype. Moreover, high secretion rates result in clear major fibre orientations across the domain: horizontal in the left and right regions, and vertical in the top and bottom regions.}
  

\subsection{Tensioning secretion and degradation in the phenotype alters invasion patterns}
\label{sec:4.4}
\begin{table}[t]
\begin{center}
\begin{tabular}{ll}
\hline
\hline
\textbf{Label } & \textbf{Collagen-related cell phenotypes}\\
\hline
\textcolor{red}{$\star^1$} & small secretion with large degradation rates \\ 
\hline
\textcolor{red}{$\star^2$} & large secretion and large degradation rates \\
\hline
\textcolor{red}{$\star^3$} & large secretion with small degradation rates \\
\hline
\hline
\end{tabular}
\end{center}
\caption{\textbf{\YY{Secretion and degradation cell phenotypes investigated in Section \ref{sec:4.4}.}}}
\label{Tab:4.4}
\end{table}

\begin{figure}[htbp]
\centering
\includegraphics[width=0.85\textwidth]{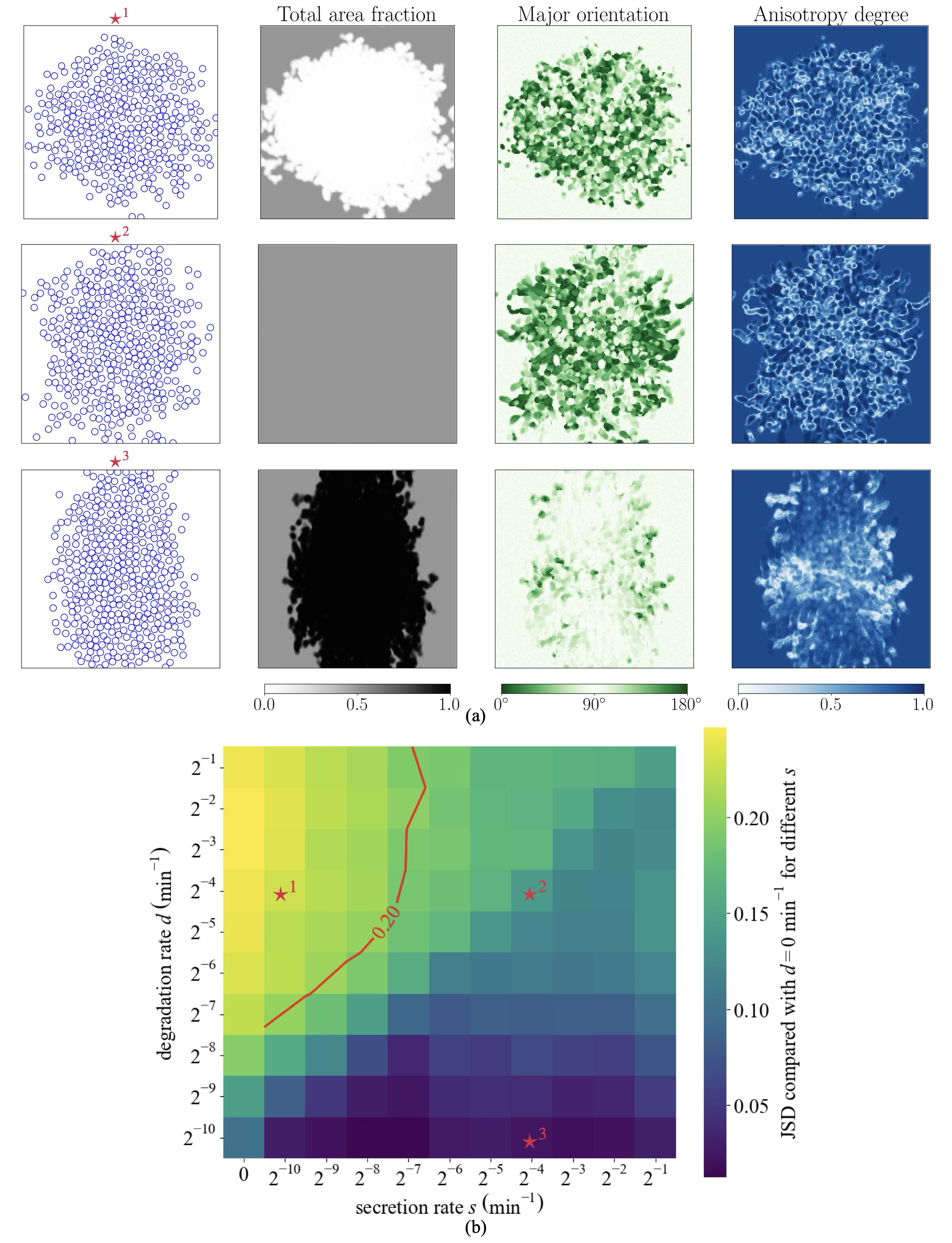}
\caption{\textbf{Tensioning secretion and degradation in the phenotype alters invasion patterns.} Results from a parameter sweep across secretion ($s$) and degradation ($d$) rates. (a) Cell and collagen fibre distributions at $42$ hours with \textcolor{red}{$\star^1$}:~$\left(s, d\right)=\left(2^{-10}, 2^{-4}\right)$ $\rm{min}^{-1}$, \textcolor{red}{$\star^2$}:~$\left(s, d\right)=\left(2^{-4}, 2^{-4}\right)$ $\rm{min}^{-1}$, and \textcolor{red}{$\star^3$}:~$\left(s, d\right)=\left(2^{-4}, 2^{-10}\right)$ $\rm{min}^{-1}$, respectively. (b) Heat map showing the similarity of the average cell density along the horizontal direction, compared to that of the non-degrading cell phenotype ($d = 0, \rm{min}^{-1}$) at 42 hours. The metric used is the Jensen-Shannon Distance (JSD), as explained in the Supplementary Information (Section~\ref{SI:JSD}): a value of $0$ indicates identical distributions, while $1$ indicates complete dissimilarity. Simulation setup: Setup2 in Section~\ref{Subsection:Pseudocodes}. The initial conditions are visualised in Figure~\ref{fig:ICs} in Supplementary Information (Section~\ref{Sec:ICS}), with $100$ cells, $\lambda_1+\lambda_2=0.5$, $a=1-\lambda_2/\lambda_1=0.9$, vertical $\hat{\mathbf{v}}_1$ and horizontal $\hat{\mathbf{v}}_2$. }
\label{Fig5}
\end{figure}

To investigate how a phenotype that both secretes and degrades fibres affects collagen fibre dynamics and cell invasiveness, we perform a parameter sweep across different fibre secretion and degradation rates. We consider a scenario in which there is very strong contact guidance initially due to a high area fraction of aligned collagen. Figure~\ref{Fig5}(a) illustrates the cell and collagen fibre distributions for small secretion with large degradation rates (\textcolor{red}{$\star^1$}), large secretion and large degradation rates (\textcolor{red}{$\star^2$}), and large secretion with small degradation rates (\textcolor{red}{$\star^3$}), \YY{see~Table \ref{Tab:4.4}}. For a cell phenotype that predominantly degrades collagen (\textcolor{red}{$\star^1$}), the initial collagen distribution is degraded \YY{and becomes isotropically distributed}, reducing the strength of contact guidance while facilitating cell invasion in all directions. This is consistent with results in Figure~\ref{Fig4}(a). For a cell phenotype that equally secretes and degrades collagen (\textcolor{red}{$\star^2$}), the initially aligned fibres are degraded: this results in reduced contact guidance so that newly secreted collagen is relatively isotropic. \YY{Comparing the \textcolor{red}{$\star^2$} phenotype with \textcolor{red}{$\star^1$}, we observe that migration is facilitated in all directions, with a stronger tendency in the vertical direction. This arises because the higher secretion rate of the \textcolor{red}{$\star^2$} phenotype produces denser collagen fibres than the \textcolor{red}{$\star^1$} phenotype. However, the large degradation rate of the \textcolor{red}{$\star^2$} phenotype subsequently leads to a weakening of these vertical cues, thereby eventually facilitating horizontal migration.} Finally, for a cell phenotype that predominantly secretes collagen (\textcolor{red}{$\star^3$}), the initial anisotropy, and thus the contact guidance cues, are reinforced in the top and bottom regions. \YY{This occurs because} cells in these regions tend to migrate vertically due to population pressure, even in the absence of contact guidance. This promotes enhanced invasion along the vertical axis, while horizontal invasion is suppressed because guidance cues from population pressure are in tension with those from the collagen distribution. This is consistent with results in Figure~\ref{Fig4}(b).


We also examine, given strong vertical contact guidance cues initially, how dominant the degradation phenotype must be for cells to effectively modulate the collagen fibre distribution and achieve isotropic invasion. To this end, we compare one-dimensional cell density profiles, generated by averaging the cell density in the $y$-direction, given different collagen degradation rates. The metric we use to compare these density profiles is the Jensen-Shannon distance (JSD), which is a bounded measure of how far apart two distributions are, with smaller values of the JSD indicating higher similarity (see Supplementary Information Section~\ref{SI:JSD}). Here, we take $\rm{JSD} = 0.2$ as the threshold beyond which invasion is substantially different from the corresponding pattern observed with the same level of secretion but no degradation. \YY{Figure~\ref{Fig5}(b)} shows that once the secretion rate exceeds a critical value ($s = 2^{-7}$ $\rm{min}^{-1}$), the cell population cannot overcome the initial vertical contact guidance cues no matter how rapid the rate of degradation.


\subsection{Showcase of invasion patterns across a range of different phenotypes}
\label{sec:4.5}


\begin{figure}[htbp]
    \centering
    \includegraphics[width=0.9\linewidth]{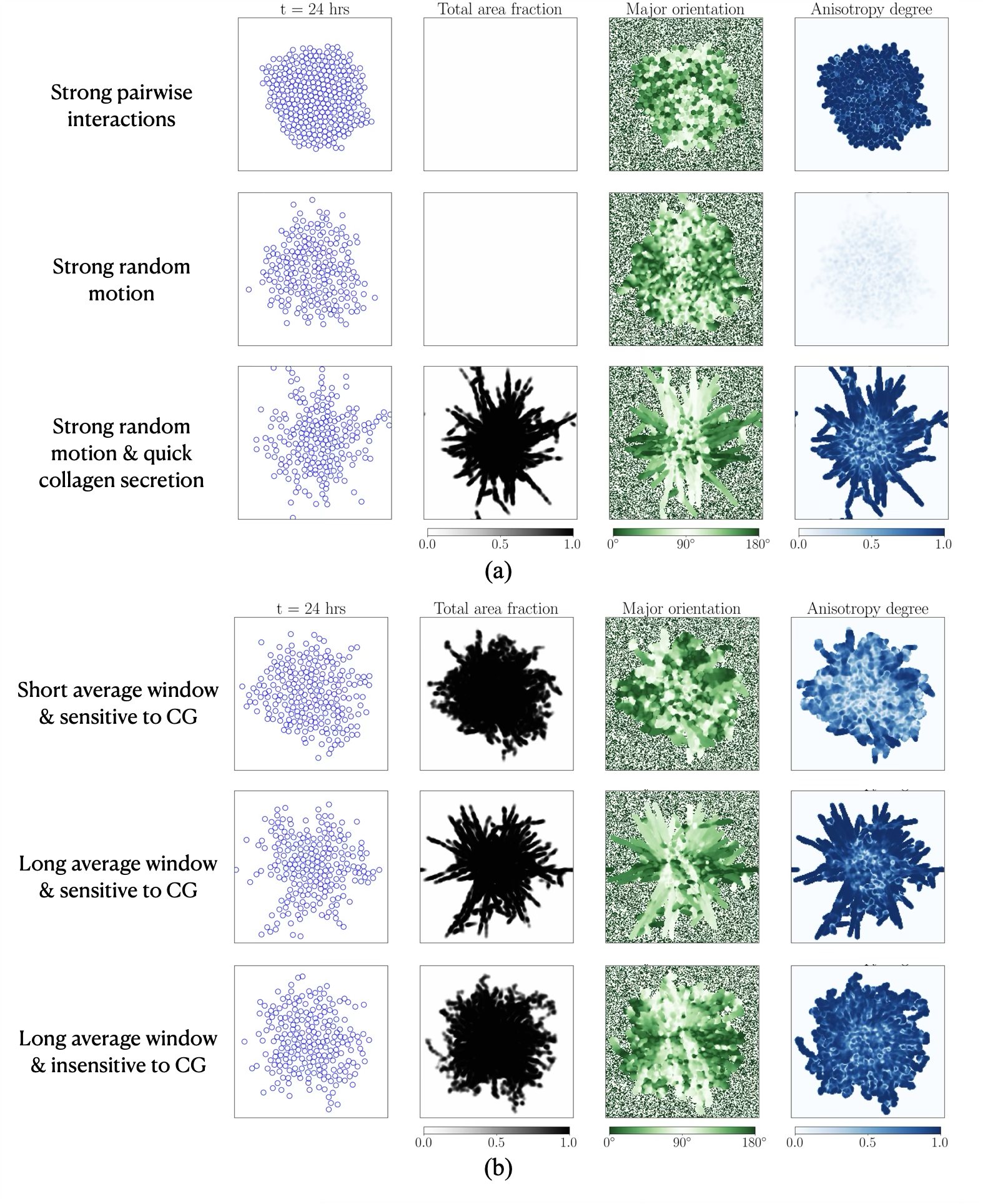}
    \caption{\textbf{Showcasing the range of invasion patterns arising from different cell phenotypes.} Cell and collagen fibre distributions at $24$ hours are shown for: (a) varying motility parameters, with $(D, \epsilon) = (0.075, 0.05)$ $\text{}\mu\text{m}^2\text{ min}^{-1}$ (top row), and $(D, \epsilon) = (0.3, 0.05)$ $\text{}\mu\text{m}^2\text{ min}^{-1}$ (middle and bottom rows); (b) varying contact guidance (CG) responses, with $(m, \bar{\lambda}) = (1\text{ min}, 0.005)$ (top row), $(m, \bar{\lambda}) = (1200\text{ min}, 0.005)$ (middle row), and $(m, \bar{\lambda}) = (1200\text{ min}, 0.9)$ (bottom row). The collagen degradation rate is fixed at $d=0.0005\text{ min}^{-1}$ for all cases, while the secretion rate is $s=0.05\text{ min}^{-1}$, except in the first two rows of (a), where $s=5.05\times 10^{-6}\text{ min}^{-1}$ to ensure that the total fibre density remains constant during invasion. Simulation setup: Setup2 in Section~\ref{Subsection:Pseudocodes}. Initial conditions are visualised in Figure~\ref{fig:ICs} in the Supplementary Information (Section~\ref{Sec:ICS}), with 100 cells, $\lambda_1 + \lambda_2 = 0.01$, and $a = 1 - \lambda_2 / \lambda_1 = 0$, corresponding to uniformly randomly distributed collagen fibres.}
    \label{Fig:6}
\end{figure}


Finally, by varying specific parameter values listed in Table~\ref{tab:ParamAndVar}, we showcase the wide \YY{range of cell} invasion and collagen fibre patterns that can arise due to different cell phenotypes. Figure~\ref{Fig:6}(a) demonstrates that when cell motility is primarily driven by pairwise interactions rather than random motion, the colony remains compact with a high degree of radial symmetry. Because the cell motility is less stochastic in nature, the secreted collagen fibres exhibit a high degree of alignment. In contrast, when cell migration is dominated by random motility, the resulting cell cluster is more dispersed, and the secreted fibres appear less aligned. Another key observation is that finger-like invasion patterns emerge only when cells secrete collagen fibres sufficiently quickly (bottom row of Figure~\ref{Fig:6}(a)). In this regime, cells at the edge of the cluster show a high degree of persistence, reinforcing both the orientation and anisotropy of the secreted fibres. This, in turn, further guides the direction of cell migration.

Figure~\ref{Fig:6}(b) shows that varying the length of the averaging window (which \YY{can be referred to as the memory window, and} determines the alignment of secreted collagen fibres) significantly influences the patterns of cell invasion. A longer averaging window \YY{(i.e.~longer memory)} smooths out the effects of random motion and results in the secretion of more aligned and spatially organised collagen fibres, which in turn enhances the emergence of a finger-like invasion pattern. On the other hand, a short averaging window \YY{(i.e.~short memory)} leads to more random motility, and less well aligned collagen fibres. Additionally, the sensitivity of cells to contact guidance cues from the underlying collagen fibres plays a crucial role in shaping the patterns of cell invasion. If cells only respond to contact guidance cues when the collagen fibre area fraction is high (i.e.~large $\bar{\lambda}$) then, despite an aligned and structured fibre field (due to the long averaging window), the cells do not exhibit finger-like behaviour. 


\section{Discussion}
\label{sec:discussion}

In this study, we developed a computationally tractable mathematical model to investigate the role of ECM-generated contact guidance in directing collective cell migration. The model incorporates biologically realistic cell--cell interactions, including volume filling and cell--cell adhesion, while focusing on the reciprocal influence between migrating cells and the orientation of collagen fibres. Through an extensive parameter sweep, we systematically examined how variations in cell phenotype impact collective invasion behaviours within the ECM. A key advantage of the representation of the ECM as a tensorial field is the ease with which we can account for its density- and anisotropy-dependent contact guidance effects on cells. 

Our model demonstrates several key findings. Firstly, when considering a region initially devoid of collagen fibres, collective cell invasion generates distinct patterns in collagen fibre deposition, with migrating cells secreting fibres in the direction of outward expansion. Cells at the leading edge of the population produce highly aligned fibres due to directed movement, while restricted motility behind the front due to a lack of free space results in a more isotropic fibre distribution. Secondly, in non-ECM-modulating phenotypes, where cells respond to but do not remodel collagen fibres, both collagen fibre area fraction and anisotropy significantly affect collective invasion patterns. A higher collagen area fraction restricts cell migration in the minor fibre direction, while increased fibre anisotropy further confines invasion along the dominant fibre orientation. These matrix properties indirectly influence cell proliferation, as denser and more aligned collagen environments lead to more compact cell clusters. Thirdly, collagen-degrading phenotypes reduce contact guidance by degrading fibres, resulting in more dispersed invasion patterns and enhanced proliferation. In contrast, collagen-secreting phenotypes create spatially varied invasion behaviours, with the direction of cell migration in the peripheral regions either more aligned with the major or minor fibre direction of the initial collagen bed, depending on their location. As a result, the cells generate complex, region-specific changes in fibre anisotropy as new collagen is deposited. Varying the balance of collagen secretion and degradation rates significantly alters invasion patterns and collagen fibre dynamics. When degradation dominates, fibres are rapidly removed, enabling uniform invasion, while high secretion rates lead to a reinforcement of existing anisotropy, directing invasion along preferred orientations. Finally, a wide range of different invasion patterns can be established through variation of the relative strengths of cell--cell interactions and random motility, as well as the extent to which cells respond to contact guidance cues.

Our model investigates the role of contact guidance in collective cell invasion using a general and modular approach, allowing for easy extensions and applications to specific biological processes. For instance, to explore wound healing, we could additionally model the dynamics of relevant chemical signals and the response of cells (e.g.~chemotaxis) to them. It is also possible to extend the model to include populations of cells of mixed phenotype, and switching of cells between different phenotypes, as observed in a range of systems (e.g.,~\cite{rognoni2018fibroblast, crossley2024phenotypic}). Recent experiments also suggest that collagen fibres can mature over time, becoming increasingly resistant to remodelling during tumour progression~\cite{fang2014collagen}. Modelling such behaviours would involve tracking the time history of the collagen fibres. Furthermore, collagen types I and III are known to have distinct roles in tissue injury and regeneration~\cite{singh2023regulation}, and we could exploit the model to track different collagen subtypes.


\section*{Acknowledgements}

The authors would like to thank Paul Riley and Chloe Stewart for very helpful discussions during development of the model. YY is funded by the Engineering and Physical Sciences Research Council (EP/W524311/1). This work was supported by a grant from the Simons Foundation (MP-SIP-00001828, REB). For the purpose of open access, the authors have applied a CC BY public copyright licence to any author accepted manuscript arising from this submission.


\renewcommand{\thesection}{S\arabic{section}}
\renewcommand{\thefigure}{S\arabic{figure}}

\setcounter{section}{0}
\setcounter{figure}{0}

\clearpage

\noindent
\textbf{\LARGE Supplementary Information}

\section{Interpretation of $\mathbf{\Omega}$ and matrix normalisation}
\label{normalisation}

\begin{figure}[h]
    \centering
    \includegraphics[width=0.8\linewidth]{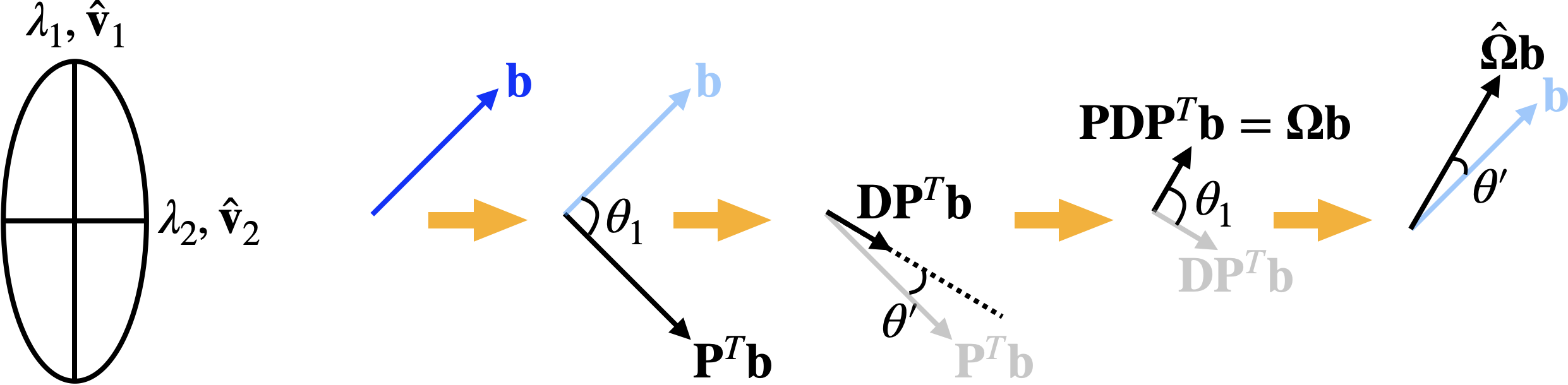}
    \caption{\textbf{Schematic diagram of $\hat{\mathbf{\Omega}}\mathbf{b}$} given $\mathbf{\Omega}=\lambda_1\hat{\mathbf{v}}_1\hat{\mathbf{v}}_1^\mathrm{T}+\lambda_2\hat{\mathbf{v}}_2\hat{\mathbf{v}}_2^\mathrm{T}$ and $\hat{\mathbf{\Omega}}$ as defined in Equation~\eqref{omega normal}.}
    \label{fig:omega_reorient}
\end{figure}

Suppose the major fibre orientation $\hat{\mathbf{v}}_1$ makes an angle $\theta_1$ with the positive $x$-axis, $\hat{\mathbf{v}}_1 = (\cos(\theta_1), \sin(\theta_1))^\mathrm{T}$. As $\hat{\mathbf{v}}_2$ is orthonormal to $\hat{\mathbf{v}}_1$, $\hat{\mathbf{v}}_2=(-\sin(\theta_1), \cos(\theta_1))^\mathrm{T}$. Based on Equation~\eqref{eq:Omega} of the main text we have
\begin{align}
\begin{split}
\mathbf{\Omega} &=\lambda_1\hat{\mathbf{v}}_1\hat{\mathbf{v}}_1^\mathrm{T} + \lambda_2\hat{\mathbf{v}}_2\hat{\mathbf{v}}_2^\mathrm{T}\\
&=
\begin{bmatrix}
| & |\\
\hat{\mathbf{v}}_1 & \hat{\mathbf{v}}_2\\
| & |
\end{bmatrix}
\begin{bmatrix}
\lambda_1 & 0\\
0 & \lambda_2
\end{bmatrix}
\begin{bmatrix}
\mbox{---} & \hat{\mathbf{v}}_1 & \mbox{---}\\
\mbox{---} & \hat{\mathbf{v}}_2 & \mbox{---}
\end{bmatrix}\\
& = \underbrace{\begin{bmatrix}
\cos(\theta_1) & -\sin(\theta_1)\\
\sin(\theta_1) & \cos(\theta_1)
\end{bmatrix}}_{:=\mathbf{P}\atop\text{anticlockwise rotation of }\theta_1}
\underbrace{\begin{bmatrix}
\lambda_1 & 0\\
0 & \lambda_2
\end{bmatrix}}_{:=\mathbf{D}\atop\text{rescale and re-orient}}
\underbrace{\begin{bmatrix}
\cos(\theta_1) & \sin(\theta_1)\\
-\sin(\theta_1) & \cos(\theta_1)
\end{bmatrix}}_{:=\mathbf{P}^\mathrm{T}\atop\text{clockwise rotation of }\theta_1}.
\end{split}   
\end{align}
For any vector $\mathbf{b}$, $\mathbf{\Omega}\mathbf{b}$: (i) rotates $\mathbf{b}$ clockwise by $\theta_1$ and obtains $\mathbf{P}^\mathrm{T}\mathbf{b}$; (ii) rescales the first and the second elements of $\mathbf{P}^\mathrm{T}\mathbf{b}$ by $\lambda_1$ and $\lambda_1$, respectively, and obtains $\mathbf{D}\mathbf{P}^\mathrm{T}\mathbf{b}$.~Unless the collagen fibres are isotropic $\lambda_1=\lambda_2$, the rescaling also leads to re-orientation of an angle $\theta'$; (iii) rotates $\mathbf{D}\mathbf{P}^\mathrm{T}\mathbf{b}$ anticlockwise by $\theta_1$ and obtains $\mathbf{P}\mathbf{D}\mathbf{P}^\mathrm{T}\mathbf{b}=\mathbf{\Omega}\mathbf{b}$. Step (ii) breaks the length-preserving property, therefore in order to obtain $\hat{\mathbf{\Omega}}$ such that $|\hat{\mathbf{\Omega}}\mathbf{b}|=|\mathbf{b}|$, we define
\begin{align}
\hat{\mathbf{\Omega}} = C_\mathbf{b}(\mathbf{\Omega})\mathbf{\Omega},\quad\text{where }
C_\mathbf{b}(\mathbf{\Omega}) = |\mathbf{b}|\big/|\mathbf{\Omega}\mathbf{b}|.
\label{omega normal}
\end{align}
Equation~\eqref{omega normal} defines the normalisation of $\mathbf{\Omega}$ and Figure~\ref{fig:omega_reorient} visualises how $\hat{\mathbf{\Omega}}$ reorients a vector $\mathbf{b}$. Similarly, to normalise $\mathbf{M}$, as defined in Equation~\eqref{eq:M_rm} of the main text, we have $\hat{\mathbf{M}} = |\mathbf{b}|/|\mathbf{M}\mathbf{b}|$ for any vector $\mathbf{b}$.


\section{The ranges for $\lambda_{1, 2}$ and $\lambda_{1}+\lambda_{2}$}
\label{lambda_ranges}

The total area fraction occupied by collagen fibres ($\lambda_1+\lambda_2$) lies between zero and one, as
\begin{equation}
\begin{split}
    \lambda_1+\lambda_2 &=\text{tr}({\mathbf{\Omega}}) \qquad \text{[as }\lambda_{1, 2}\text{ are eigenvalues]}\\
&=\text{t}r\left(\frac{1}{\pi}\int_{0}^{\pi}\hat{\mathbf{u}}(\theta)\hat{\mathbf{u}}^\mathrm{T}(\theta)\rho(\theta, \mathbf{x}, t) \mathrm{d} \theta\right)\qquad\text{ [by Equation~\eqref{eq:GeneralRepOfFibre}]}\\
&=\frac{1}{\pi}\int_{0}^{\pi}tr\left(\hat{\mathbf{u}}(\theta)\hat{\mathbf{u}}^\mathrm{T}(\theta)\right)\rho(\theta, \mathbf{x}, t) \mathrm{d} \theta\\
&=\frac{1}{\pi}\int_{0}^\pi \rho(\theta, \mathbf{x}, t) \qquad\text{ [as }\hat{\mathbf{u}}\text{ is a unit vector]}\\
&\in [0, 1]\qquad\text{ [by the definition of }\rho\in [0, 1]\text{].}
\end{split}
\end{equation}
Moreover, as $\mathbf{\Omega}$ is a symmetric positive semi-definite matrix (see Equation~\eqref{eq:Omega}), $\lambda_{1, 2}\in [0, 1]$. 


\section{Average number of repulsive neighbours per cell}

\begin{figure}[tbh!]
    \centering
    \includegraphics[width=1.0\linewidth]{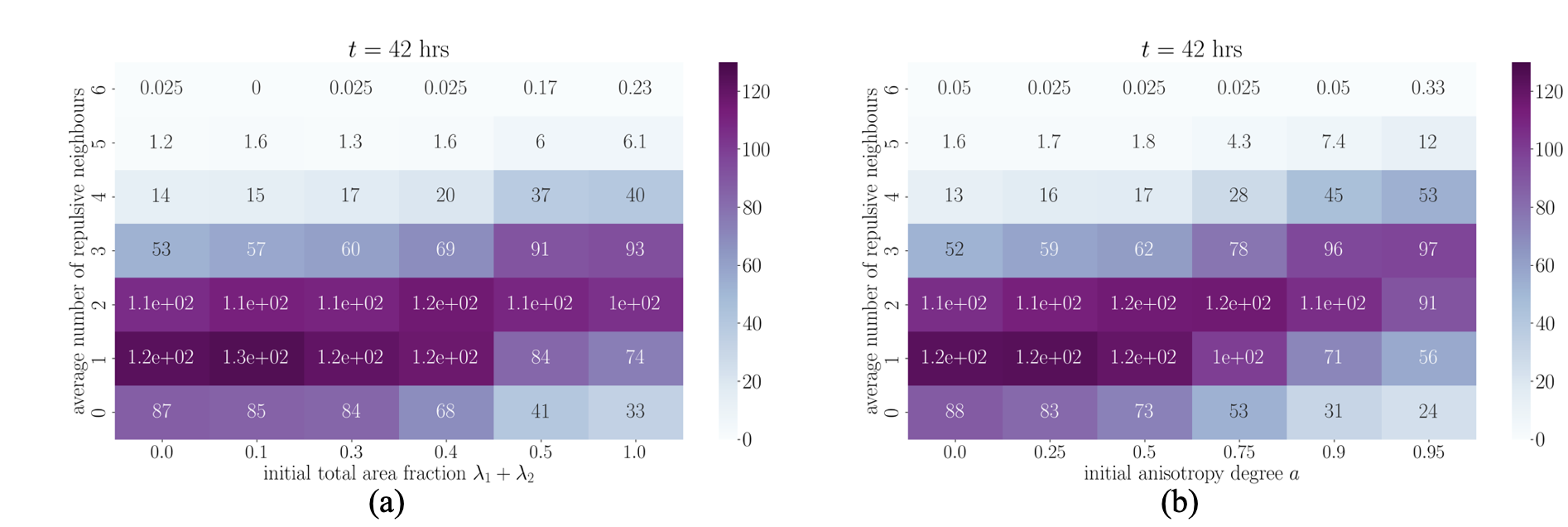}
    \caption{\textbf{Average number of repulsive neighbours per cell} given no secretion and degradation. Results are averaged over $40$ simulations, given different total area fractions ($\lambda_1 + \lambda_2$) and anisotropy degrees ($a=1-\lambda_2/\lambda_1$) of collagen fibres.}
    \label{fig:CG-ONLY_supp}
\end{figure}

Figure~\ref{fig:CG-ONLY_supp} shows the average number of cells within the repulsive range of $2^{1/6}\sigma$ of a given cell, where $\sigma$ represents the cell diameter. We observe that denser and more anisotropic collagen fibres result in more compact cell clusters, causing each cell to have more repulsive neighbours, thus less space to proliferate.


\section{Jensen-Shannon distance (JSD)}
\label{SI:JSD}

To quantify the similarity between two distributions, we use the Jensen-Shannon distance (JSD).~The JSD between two probability vectors $\mathbf{p}$ and $\mathbf{q}$ is defined as
\begin{align}
    \mathrm{JSD}(\mathbf{p}, \mathbf{q}):=\sqrt{\frac{D(\mathbf{p} \parallel \mathbf{m}) + D(\mathbf{q} \parallel \mathbf{m})}{2}},
\end{align}
where $\mathbf{m}$ is the point-wise mean of $\mathbf{p}$ and $\mathbf{q}$, and $D(A \parallel B)=\sum_{x\in \mathcal X}A(x)\log\left({A(x)}/{B(x)}\right)$ is the Kullback-Leibler divergence. The JSD is symmetric and bounded between $[0, 1]$, and the similarity between two distributions is greater for JSD closer to zero, and vice versa.


\section{Initial conditions for simulations}
\label{Sec:ICS}

\begin{figure}[ht!]
    \centering
    \includegraphics[width=1.0\linewidth]{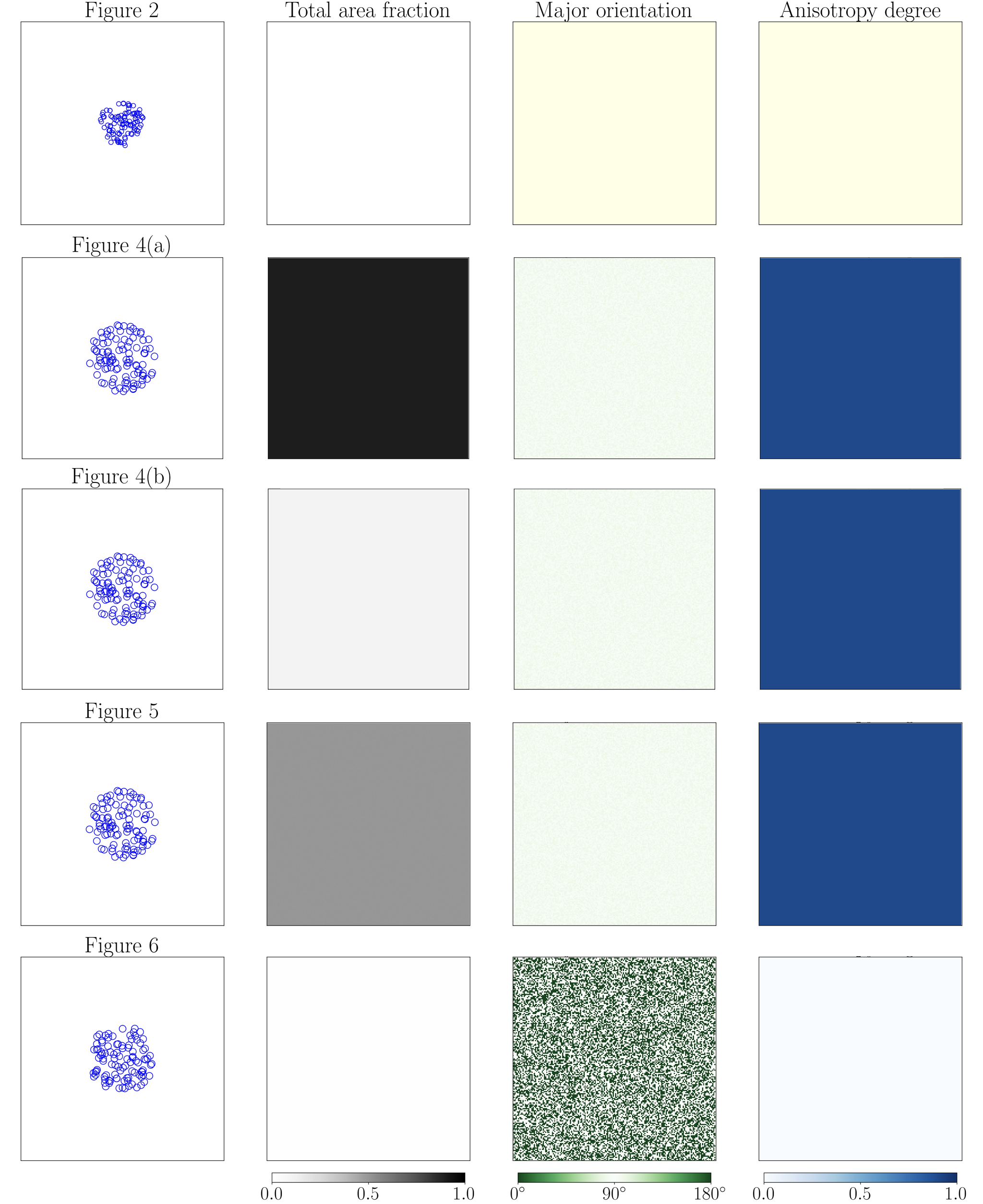}
    \caption{\textbf{Initial conditions for results in the main text.} \YY{$100$ cells are initially arranged in a densely packed circular disc at the centre of the domain with radius $60\, \mu$m, forming a confluent configuration.} Pale yellow backgrounds indicate regions devoid of collagen fibres. Simulation details: Setup1 with a larger domain and longer simulation time is used for Figure~\ref{Fig2}, while Setup2 is used for Figures~\ref{Fig3}–\ref{Fig:6}.}
    \label{fig:ICs}
\end{figure}

In all simulations, $100$ cells are initially arranged in a densely packed circular disc at the centre of the domain \YY{with radius $60\, \mu$m}, forming a confluent configuration. For Figure~\ref{Fig2} in the main text, we initialise the system without any collagen fibres. For Figure~\ref{Fig3}(a), (c), and (e), the collagen fibres are initialised with a high anisotropy degree ($a = 0.9$), while the total fibre density ($\lambda_1 + \lambda_2$) is varied among $0.1$, $0.4$, and $1$. For Figure~\ref{Fig3}(b), (d), and (f), the total fibre density is fixed at $\lambda_1 + \lambda_2 = 0.8$, while the anisotropy degree is varied among $0.0$, $0.5$, and $0.95$. In Figure~\ref{Fig4}(a), collagen fibres are initialised with a high density ($\lambda_1 + \lambda_2 = 0.9$), whereas in Figure~\ref{Fig4}(b), a low density ($\lambda_1 + \lambda_2 = 0.1$) is used. In both cases, the fibres are aligned ($a = 0.9$). For Figure~\ref{Fig5}, the collagen fibres are initialised with $\lambda_1 + \lambda_2 = 0.5$ and $a = 0.9$. In Figure~\ref{Fig:6}, a sparse and isotropic collagen distribution is used, with $\lambda_1 + \lambda_2 = 0.01$ and $a = 0$. For Figures~\ref{Fig2}–\ref{Fig5}, the angle between the major fibre orientation and the positive $x$-axis, denoted $\angle \hat{\mathbf{v}}_1$, is sampled from the interval $\pi/2 + U[-\pi/18, \pi/18]$. In contrast, for Figure~\ref{Fig:6}, the collagen fibres are randomly oriented, with $\angle \hat{\mathbf{v}}_1$ sampled uniformly from $U[-\pi, \pi]$. All initial conditions, except for those in Figure~\ref{Fig3}, are visualised in Figure~\ref{fig:ICs}.


\section{\YY{Different collagen modulating kernels $\omega$ lead to similar dynamics}}

\begin{figure}[]
\centering
    \includegraphics[width=0.9\linewidth]{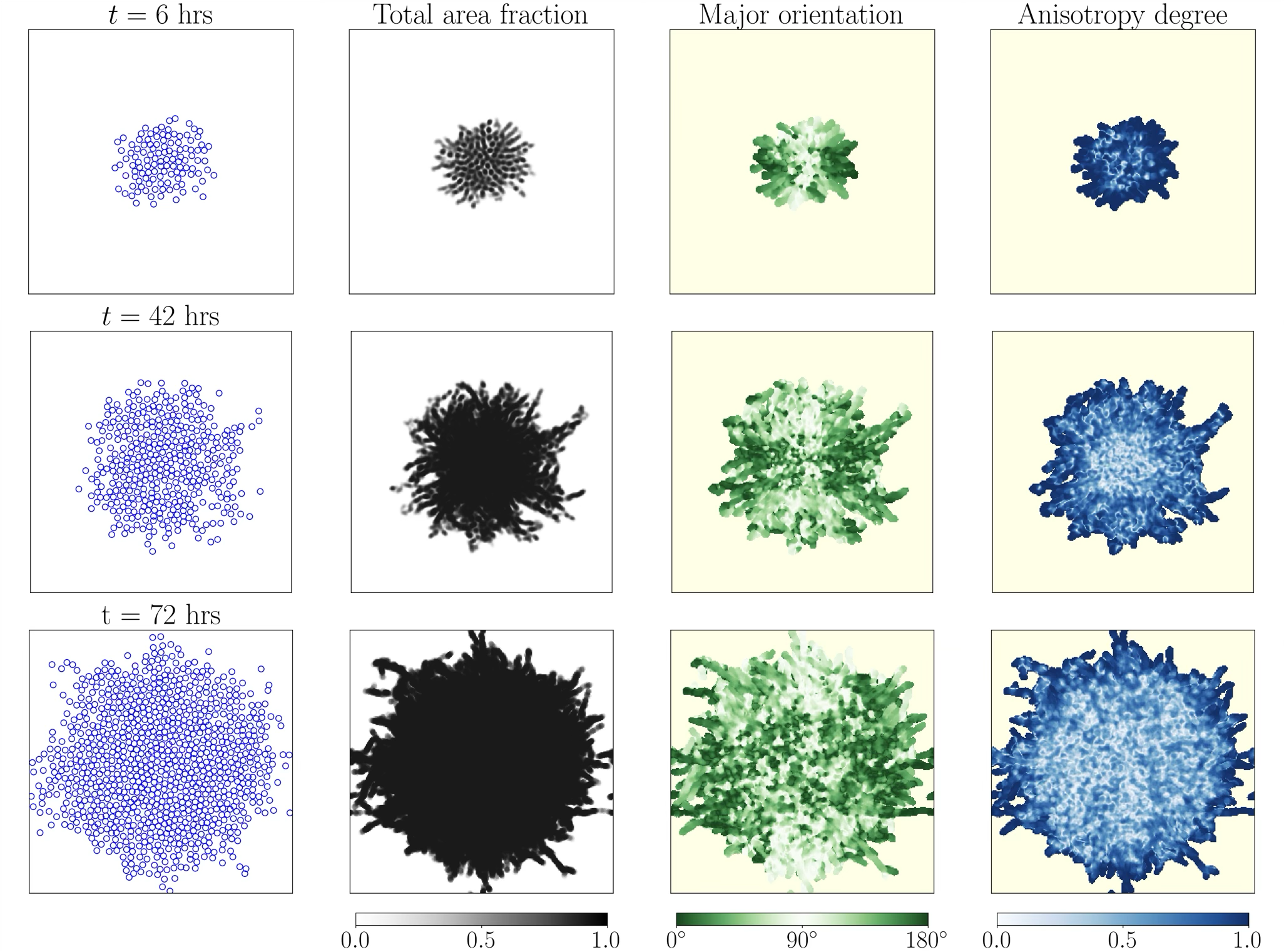}
    \caption{{\YY{\textbf{Cell migration patterns are shown when the collagen-modulating kernel $\omega$ is taken as a Heaviside function with support $\sigma/2$, corresponding to the cell radius.} This figure can be compared with Figure~\ref{Fig2} in the main text.}}}
\label{Fig:heaviside_w}
\end{figure}

\begin{figure}[tbh!]
    \centering
    \includegraphics[width=1.0\linewidth]{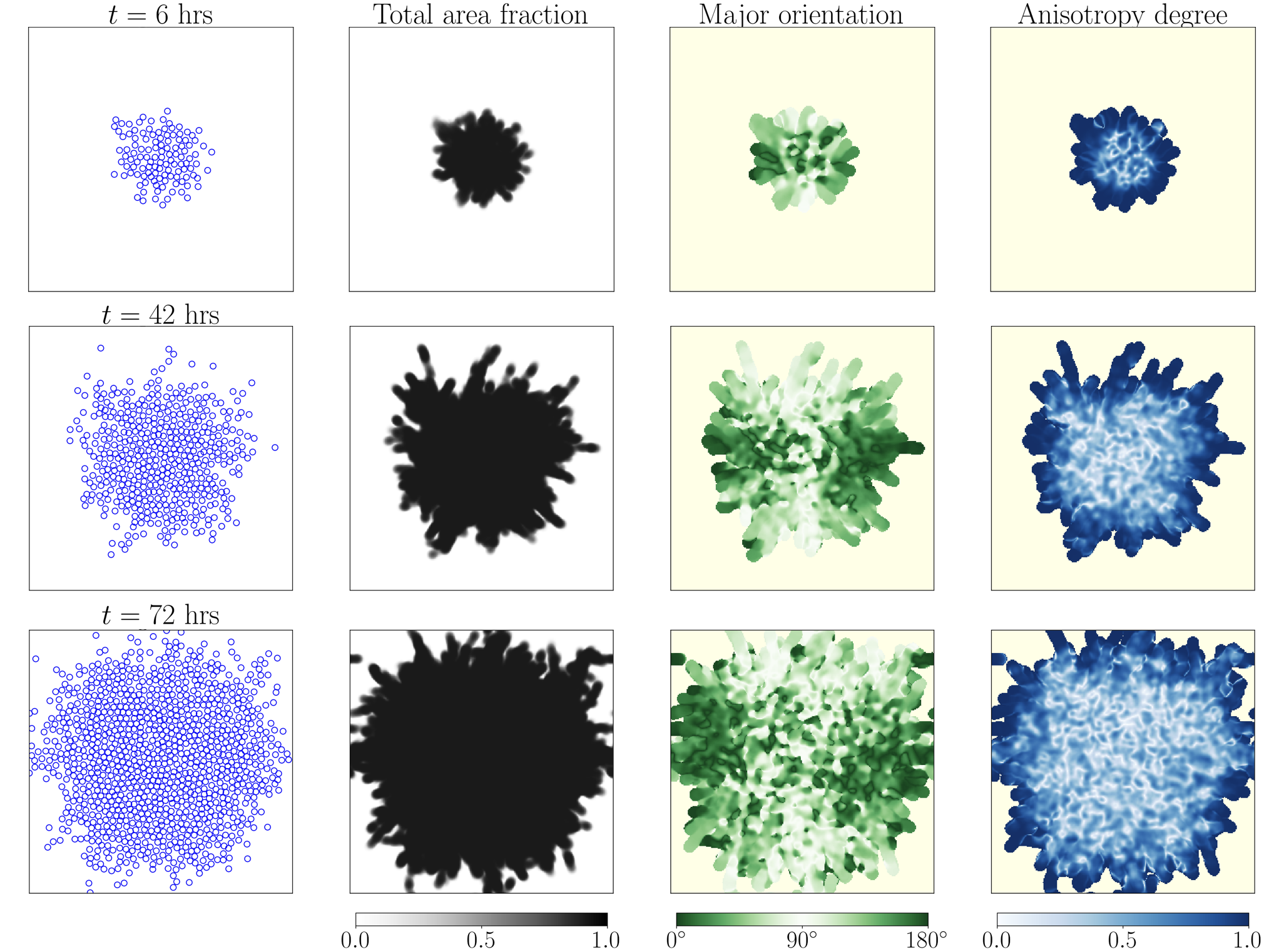}
    \caption{\YY{\textbf{Cell migration patterns are shown when the range of collagen secretion and degradation extends beyond the space occupied by a single cell; that is, the range of collagen modulation is $\sigma$, with $\sigma/2$ representing the cell radius. }This figure can be compared with Figure~\ref{Fig2} in the main text.}}
    \label{fig:nonlocal_w}
\end{figure}

\YY{Recall that the collagen-modulating kernel $\omega$, defined in Equation~(\ref{eq:Omega}), characterises the strength and range of collagen degradation and secretion. We now investigate whether the cell and collagen fibre dynamics are sensitive to the functional form of $\omega$. Figure~\ref{Fig:heaviside_w} illustrates cell migration and collagen fibre dynamics when $\omega$ is taken to be a Heaviside function:
\begin{equation}
    \label{eq:heaviside}
    \omega\left(\mathbf{X}^i, \mathbf{x},\sigma\right) = 
    \begin{cases}
    1, & \quad |\mathbf{X}^i-\mathbf{x}|\leq \sigma/2,\\
    0, & \quad |\mathbf{X}^i-\mathbf{x}|> \sigma/2.
    \end{cases}
\end{equation}
We note that since $\sigma/2$ is a measure of the cell radius, this functional form is intended to restrict secretion and degradation locally to a cell. On the other hand, Figure~\ref{fig:nonlocal_w} shows the cell migration and collagen fibre dynamics for
\begin{equation}
    \label{eq:cell-fibreWeightFunction}
    \omega\left(\mathbf{X}^i, \mathbf{x},\sigma\right) = 
    \begin{cases}
    1-\frac{|\mathbf{X}^i-\mathbf{x}|}{\sigma}, & \quad |\mathbf{X}^i-\mathbf{x}|\leq \sigma,\\
    0, & \quad |\mathbf{X}^i-\mathbf{x}|> \sigma.
    \end{cases}
\end{equation}
In this case the range over which collagen is secreted and degraded is doubled to $\sigma$. Comparing Figures~\ref{Fig:heaviside_w}--\ref{fig:nonlocal_w} with Figure~\ref{Fig2}, we observe no qualitative differences in cell or collagen fibre dynamics.}

\section{Collagen degrading phenotypes display different invasion patterns}
\label{sec: fig4_SI}

\begin{figure}[htbp]
    \centering
    \includegraphics[width=1.0\linewidth]{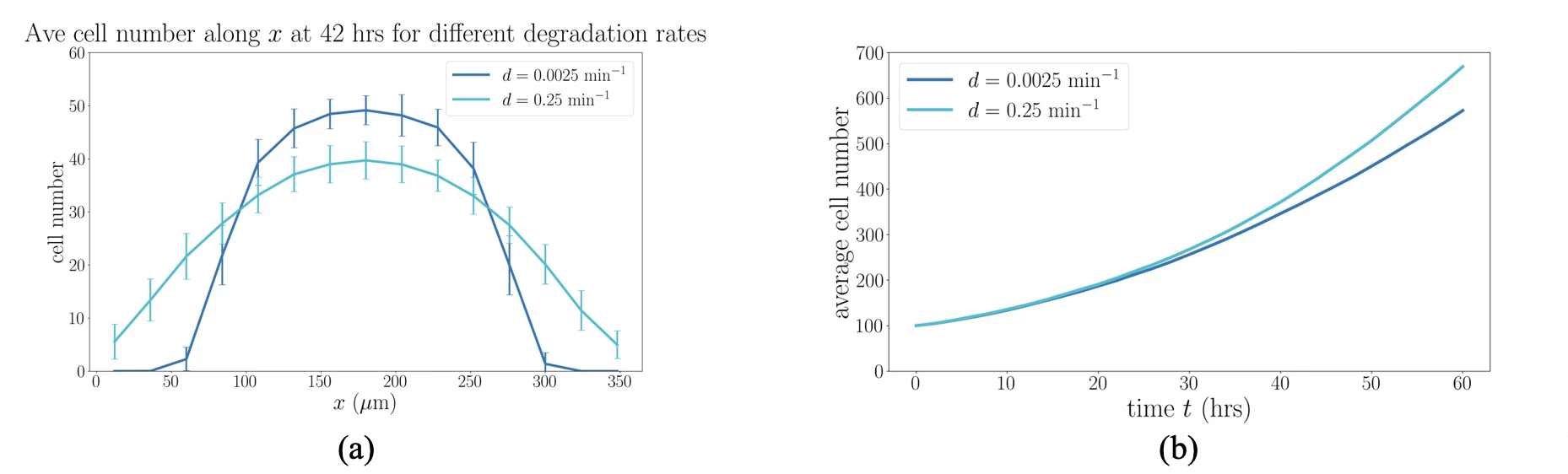}
    \caption{\textbf{Collagen degrading phenotypes enhance horizontal invasion.} (a) One-dimensional cell density profiles obtained by counting the number of cells along the $x$-direction and averaging over the $y$-direction for $d=0.0025$ $\mathrm{min}^{-1}$ and $d=0.25$ $\mathrm{min}^{-1}$ at $42$ hours. Error bars represent the standard deviation over $40$ repetitions. (b) Average cell number over time for $d=0.0025$ $\mathrm{min}^{-1}$ and $d=0.25$ $\mathrm{min}^{-1}$. This figure is related to Figure~\ref{Fig4}(a) in the main text.}
    \label{fig:fig4_SI}
\end{figure}

Figure~\ref{fig:fig4_SI}(a) compares the one-dimensional cell density profiles averaged along the $y$-direction, for slow ($d = 0.0025$ $\mathrm{min}^{-1}$) and fast ($d = 0.25$ $\mathrm{min}^{-1}$) collagen degradation. The results show that dense, vertically aligned collagen fibres restrict horizontal cell invasion, leading to a more compact, vertically striped cell pattern. Consequently, cell density-dependent proliferation is more strongly suppressed in the case of fast degradation ($d = 0.25$ $\mathrm{min}^{-1}$), as shown in Figure~\ref{fig:fig4_SI}(b).


\clearpage
\bibliographystyle{ieeetr}
\bibliography{sample}


\end{document}